\documentclass[]{aa}  

\usepackage{graphicx}
\usepackage{txfonts}
\usepackage{float}
\usepackage{enumitem}
\usepackage{ctable}
\usepackage[rightcaption]{sidecap}
\usepackage{placeins}
\begin{document} 

   \title{PRODIGE - Planet-forming disks in Taurus with NOEMA}
  \subtitle{II. Modeling the CO~(2-1) isotopologue emission of the Class~II T~Tauri disks in Taurus.}
   \author{R. Franceschi\inst{1}
        \and Th. Henning\inst{1}
        \and G. V. Smirnov-Pinchukov\inst{1}
        \and D. A. Semenov\inst{1,2}
        \and K. Schwarz\inst{1}
        \and A. Dutrey\inst{3,4}
        \and E. Chapillon\inst{5}
        \and U. Gorti\inst{6,7}
        \and S. Guilloteau\inst{3,4}
        \and V. Pi\'etu \inst{5}
        \and S. van Terwisga\inst{1}
        \and L. Bouscasse \inst{5}
        \and P. Caselli \inst{8}
        % \and C. Ceccarelli \inst{7}
        % \and N. Cunningham \inst{7}
        % \and A. Fuente \inst{8}
        \and G. Gieser \inst{8}
        \and T.-H. Hsieh \inst{8}
        \and A. Lopez-Sepulcre \inst{5,9}
        \and D. M. Segura-Cox \inst{10,8}\thanks{NSF Astronomy and Astrophysics Postdoctoral Fellow}
        \and J. E. Pineda \inst{8}
        \and M. J. Maureira  \inst{8}
        % \and Th. M\"{o}ller \inst{10}
        % \and M. Tafalla \inst{8}
        \and M. T. Valdivia-Mena \inst{8}}

   \institute{Max-Planck-Institut f\"ur Astronomie (MPIA), K\"onigstuhl 17,  69117 Heidelberg, Germany
              \and Department of Chemistry, Ludwig Maximilian University, Butenandtstr. 5-13, D-81377 Munich, Germany
              \and LAB, Universit\'e de Bordeaux, B18N, All\'ee Geoffroy, Saint-Hilaire, CS 50023, 33615 Pessac Cedex
              \and CNRS, Universit\'e de Bordeaux, B18N, All\'ee Geoffroy, Saint-Hilaire, CS 50023, 33615 Pessac Cedex
              \and IRAM, 300 Rue de la Piscine, F-38046 Saint Martin d'H\`eres, France
              \and NASA Ames Research Center, Moffett Field, CA 94035, USA
              \and Carl Sagan Center, SETI Institute, Mountain View, CA 94043, USA
              \and Max-Planck-Institut f\"{u}r extraterrestrische Physik (MPE), Gie{\ss}enbachstr. 1, D-85741 Garching bei M\"{u}nchen, Germany
              \and IPAG, Universit\'{e} Grenoble Alpes, CNRS, F-38000 Grenoble, France
              % \and Observatorio Astron\'{o}mico Nacional (IGN), Alfonso XII 3, E-28014, Madrid, Spain
              \and Department of Astronomy, The University of Texas at Austin, 2500 Speedway, Austin, TX, 78712, USA
              % \and I. Physikalisches Institut, Universität zu K\"{o}ln, Z\"{u}lpicher Str. 77, 50937 K\"{o}ln, Germany\\
              \email{franceschi@mpia.de}
              }

% \abstract{}{}{}{}{} 
% 5 {} token are mandatory
 
  \abstract
   {To understand how planets form in protoplanetary disks, it is necessary to characterize their gas and dust distribution and masses. This requires a combination of high-resolution dust continuum and molecular line interferometric observations, coupled with advanced theoretical models of protoplanetary disk physics, chemical composition, and radiative transfer.} 
   {We aim to constrain the gas density and temperature distributions as well as gas masses in several T~Tauri protoplanetary disks located in Taurus. We use the $^{12}$CO, $^{13}$CO, and C$^{18}$O~(2-1) isotopologue emission observed at $0.9\arcsec$ with the IRAM NOrthern Extended Millimeter Array (NOEMA) as part of the MPG-IRAM Observatory Program PRODIGE (PROtostars and DIsks: Global Evolution \, PIs: P.~Caselli \& Th.~Henning). Our sample consists of Class~II disks with no evidence of strong radial substructures. We use these data to constrain the thermal and chemical structure of these disks through theoretical models for gas emission.}
   {To fit the combined optically thick and thin CO line data in Fourier space, we developed the DiskCheF code, which includes the parameterized disk physical structure, machine-learning (ML) accelerated chemistry, and the RADMC-3D line radiative transfer module. A key novelty of DiskCheF is the fast and feasible ML-based chemistry trained on the extended grid of the disk physical-chemical models precomputed with the ANDES2 code. This ML approach allows complex chemical kinetics models to be included in a time-consuming disk fitting without the need to run a chemical code.}
   {We present a novel approach to incorporate chemistry into disk modeling without the need to explicitly calculate a chemical network every time. Using this new disk modeling tool, we successfully fit the $^{12}$CO, $^{13}$CO, and C${18}$O~(2-1) data from the CI, CY, DL, DM, DN, and IQ~Tau disks. The combination of optically thin and optically thick CO lines allows us to simultaneously constrain the disk temperature and mass distribution, and derive the CO-based gas masses. The best-fit disk gas masses range between 0.005 and $0.04~M_{\sun}$. These values are in reasonable agreement with the disk dust masses rescaled by a factor of 100 as well as with other indirect gas measurements via, for example, modeling of the wavelength dependence of the dust continuum emission radii, and HD and CO isotopologue emission.}
   {}
   
  %We build a disk physical and chemical model based on multiple CO isotopologs emission observed by the PRODIGE collaboration.

    
   \authorrunning{R. Franceschi \inst{1}, Th. Henning\inst{1}, D.A. Semenov\inst{1,2}, et al.}%G.V. Smirnov-Pinchukov\inst{1}, S. van Terwisga\inst{1}}

   % \institute{Max-Planck-Institut f\"ur Astronomie (MPIA),
   %            K\"onigstuhl 17,  69117 Heidelberg, Germany
   %            \and University Observatory, Faculty of Physics, Ludwig Maximilians University, Scheinerstr. 1, 81679 Munich, Germany
   %            \and Department of Chemistry, Ludwig Maximilian University, Butenandtstr. 5-13, D-81377 Munich, Germany\\
   %            \email{franceschi@mpia.de}
   %            }
   
   \titlerunning{PRODIGE: modeling the CO isotopologs emission in the Taurus region}
   \authorrunning{Franceschi et al.}
   \maketitle

   \newcommand{\molhyd}{$\mathrm{H_2} \,$}
   \newcommand{\CO}[2]{$\mathrm{^{#1}C^{#2}O}$}
%
%-------------------------------------------------------------------

\section{Introduction}
\label{sec:intro}
The total gas mass and gas density and temperature distributions are the fundamental parameters with which to understand the evolution of protoplanetary disks and the planet formation process \citep[e.g.,][]{Mordasini12}. Probing the disk gas structure is a challenging task, as most of the mass is carried by \molhyd molecules, which do not emit under the physical conditions of the outer disk regions probed by the far-infrared and radio observations. Therefore, the gas disk thermal and density distributions are studied through the emission of other, much less abundant molecules, in combination with theoretical models. The most common gas tracer is the CO molecule, as it is relatively abundant and emits at cold temperatures ($\sim$10-50~K), typical of the gas in the outer disk regions \citep[e.g.,][]{Aiwaka02, Williams14, Zhang21}. A variety of the CO isotopologues with abundances that can differ by factors of $\gtrsim 70-150\,000$, and, consequently, with emission lines of very different optical depths, enable one to trace the disk gas through a broad range of vertical heights in the molecular layer. CO molecules are the key carriers of the elemental carbon and oxygen, have low freeze-out temperatures, and are crucial to the chemical processes of gas-phase carbon and oxygen in the outer disk leading to the formation of more complex organic molecules \citep{Walsh14,Favre_ea18a,Lee_ea19}.

CO is a chemically stable molecule with a simple and well-known chemistry. The CO rotational transitions come from the so-called disk molecular layer, which extends above the disk cold midplane where it is removed from the gas phase by freeze-out, until the photodissociation region where CO is destroyed by the external or stellar radiation field \citep{Aiwaka02, DDG03, Molyarova17}. Consequently, the temperature of the CO-emitting gas in disks should be higher than about 20~K; otherwise, CO molecules would be removed from the gas phase by freeze-out \citep{Harsono15, Schwarz16, Pinte18}.

However, some recent observational studies have questioned the reliability of CO as a gas tracer in disks due to stronger CO depletion being predicted by contemporary models, leading to a much lower CO/H$_2$ ratio than the canonical interstellar medium (ISM) value of $\sim 10^{-4}$ \citep[e.g.,][]{Aikawa97, Favre_ea13_TWHya_CO, Miotello_ea18a, MAPS20_Schwarz_ea21}. Recent surveys in the Chamaeleon and Lupus star formation regions found that weak CO emission could be common in protoplanetary disks \citep{Ansdell16, Long17, Miotello17}. These studies indicate that the CO abundance can be lower by up to two orders of magnitude than the ISM value of $\sim 10^{-4}$.

There are several processes that can affect the CO abundance in the gas phase, making the conversion from CO abundance to \molhyd abundance uncertain. First, CO molecules can be frozen out in the cold disk midplane, or photodissociated in the upper disk layers by the far-ultraviolet (FUV) photons from the central star or by external FUV radiation \citep{Miotello16}. Second, chemical processes can alter the CO abundance, for example by transforming it into other molecules such as CO$_2$ and complex organics \citep[e.g.,][]{Bruderer12, Bruderer13, Schwarz18}. The photodissociation or freeze-out results in overall lower C/H and O/H abundances in the gas, which are important parameters for disk chemistry and which can hence affect the abundances of other molecular disk tracers. %further affecting the CO abundance. 
Constraining the total CO depletion factor from the line observations is a challenging task, as it requires a good understanding of the underlying chemistry as well as dust and gas dynamics and properties \citep{Krijt18, Krijt20}, but also optical depth effects and line excitation conditions, since even rare CO isotopologues can be optically thick in the dense inner disk within the CO snowline region. 

Isotopologue-selective processes need to also be considered, such as self-shielding against photodissociation or low-temperature $^{12}$C/$^{13}$C-fractionation \citep[e.g.,][]{Woods_Willacy08, Visser09}. Disk gas mass estimates based on the CO isotopologue lines could be underestimated by up to two orders of magnitude if these processes are not fully taken into account \citep{Miotello22}. These processes are accounted for in many contemporary physical-chemical models, but not all \citep[e.g.,][]{Miotello14, Miotello16, Ruaud22}. Other recent studies compared Class~I and Class~II disks and found that the CO gas-phase abundances decrease rapidly in older disks \citep[e.g.,][]{Zhang20}. Thus, to use the CO emission as a reliable proxy of the disk gas masses, we need to employ more feasible disk physical-chemical models, as well as parameter space sampling algorithms. 

In this paper, we model the millimeter emission of the optically thick and thin CO~(2-1) isotopologue lines observed in several Class~II T~Tauri disks in the Taurus star formation region as part of the MPG-IRAM large guaranteed time project PRODIGE (PROtostars to DIsks: Global Evolution; PIs: P.~Caselli \& Th.~Henning); see Sect.~\ref{sec:observations}. Previous works using the PRODIGE data are the studies of the Class 0/I protostars and streamers in the Perseus region \citep{Hsieh22, Valdivia-Mena22}. An overview of the CO observations for all targeted Class~II sources and the first analysis and modeling of these data are presented in \citet{PRODIGE-I}. To fit the CO NOEMA data, we developed a novel disk fitting tool, \texttt{DiskCheF}\footnote{\url{https://gitlab.com/SmirnGreg/diskchef/}} (Disk Chemical Fitter), which uses a parametric disk physical model, ML-accelerated gas-grain chemistry, and line radiative transfer based on the RADMC-3D code (Sect.~\ref{sec:model}). Applying DiskCheF to our CO data, we infer the best-fit disk physical structures and the corresponding gaseous masses (Sect.~\ref{sec: results}). We discuss the results and limitations of our DiskCheF fitting and compare it with previous similar studies of disk masses in Sect.~\ref{sec:diss}. %Final conclusions follow.

Recently, an ALMA study of molecular line emission in five protoplanetary disks was carried out by the MAPS collaboration \citep[(PI: K.~Oeberg, ][see the following papers]{Oberg21}, with a different science goal than the one presented in this work. While the five MAPS sources comprise both T~Tauri and Herbig~Ae stars from different star-forming regions, the disks in the PRODIGE survey are all T~Tauri systems and located in the same Taurus star-forming region.

\section{Observations}
\label{sec:observations}
We chose to perform the CO line fitting on six of the eight Class~II disks observed in our sample; namely, CI, CY, DL, DM, DN, and IQ~Tau (see Tab.\ref{table:obs_prop}). The selected systems are isolated T~Tauri stars ($M_{\star} < 1 \; M_\odot$) with ages between 1 and 4~Myr (see \citealt{PRODIGE-I} and references within), surrounded by extended disks ($R_\mathrm{CO} \gtrsim 200$~au) without strong substructures visible in the ALMA continuum observations \citep{Long18, Long19}. The two systems removed from the modeling are DG~Tau and UZ~Tau~E. The DG~Tau system is young and still embedded in an extended envelope with a jet and an outflow, and would require more dedicated efforts and a detailed disk-envelope-outflow model to reproduce complex CO data. The UZ~Tau~E system is excluded because it is a part of a binary system, and shows a large-scale gaseous arc-like structure connecting the two circumstellar disks, likely caused by their gravitational interactions. 

In this work, we focus on the CO, $^{13}$CO, and C$^{18}$O $J = 2 - 1$ emission lines detected in the framework of the PRODIGE program. 
The spectral setup, observing details, data reduction, self-calibration, and imaging are described in the Class~II PRODIGE paper by \citet{PRODIGE-I}. Briefly, NOEMA/PolyFiX observations with ten antennas in C configuration ($0.9\arcsec$ angular resolution) were taken in Band~3 at 1.3~mm in 2020, using 62.5kHz spectral chunks to spectrally resolve specific emission lines. We targeted the lines of $^{12}$CO, $^{13}$CO, C$^{18}$O, para-H$_2$CO, DCO$^+$, DCN, DNC, $^{13}$CN, cyclic C$_3$H$_2$, C$_2$D, HC$_3$N, N$_2$D$^+$, and other rare species. Each source was observed independently (no track-sharing), using one to three tracks. The average on-source integration time was 5.5~hours per disk, and the resulting $1\sigma$ rms noise is $3.7-9.2$~mJy at 0.3~km\,s$^{-1}$ resolution. 

A preliminary analysis of the combined CO~(2-1) data in \citet{PRODIGE-I} shows that a foreground cloud partially obscures the optically thick $^{12}$CO~(2-1) emission in the 4.5-6~km/s velocity range in the five disks, except for DM~Tau, DN~Tau, and IQ~Tau. The $^{12}$CO~(2-1) peak brightness temperatures range between $6.5 - 16.0$~K, while for more optically thin $^{13}$CO~(2-1) and C$^{18}$O~(2-1) these peak brightness temperatures are $1.2 - 7.5$~K and $0.2 - 1.8$~K, respectively. The moderate spatial resolution of our NOEMA observations ($100 - 150$~au) leads to smooth CO radial intensity profiles as reconstructed by the Kepler deprojection and azimuthal averaging, with no evidence of substructures. The apparent inner emission ``holes'' in the CO intensities at $r \lesssim 100-150$~au are due to beam dilution. In Fig.\ref{fig: mom0} we show the moment zero emission maps of our selected sources.

{
\sidecaptionvpos{figure}{c}
\begin{SCfigure*}
    \centering
        \includegraphics[width=0.75\textwidth]{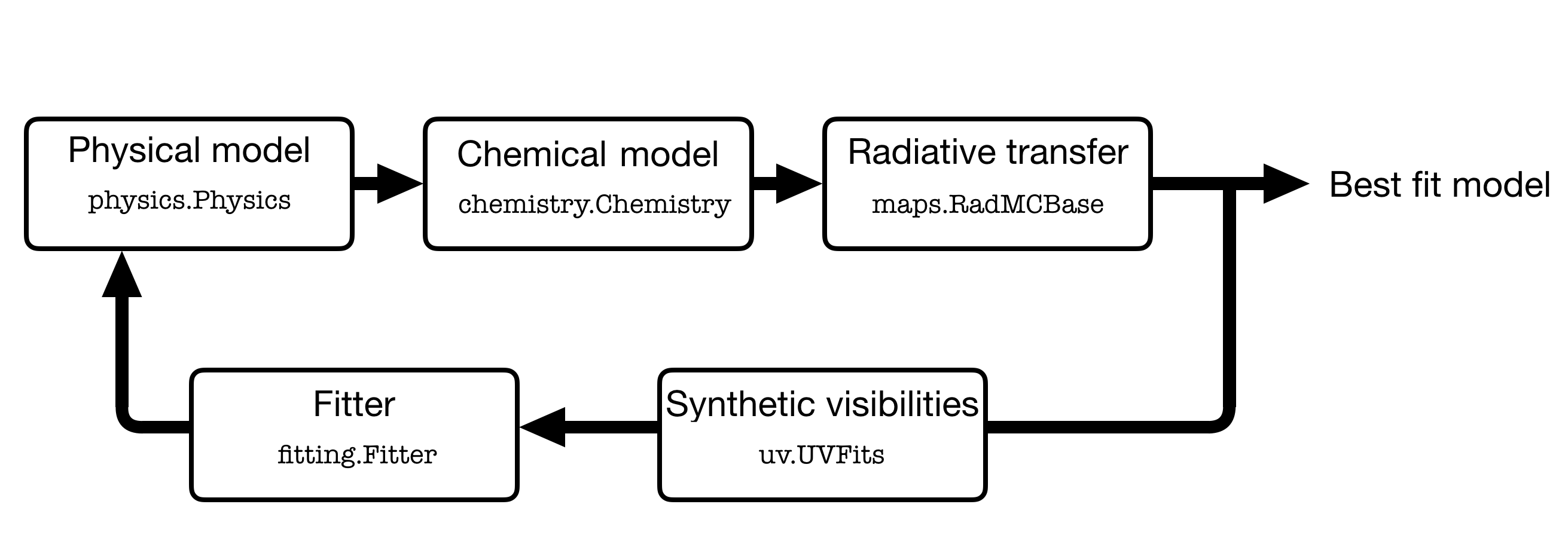}
            \caption{
            Outline of the steps in a \texttt{DiskCheF} model. For each step, we indicate which template can be used to create a new custom model.
                  }
             \label{fig:workflow}
\end{SCfigure*}
}

%A similar study of the CO isotopologue emission lines was performed by the MAPS ALMA large program \citep[(PI: K.~Oeberg, ][see the following papers]{Oberg21}, with a different science goal than the one presented in this work. While the five MAPS sources comprise both T~Tauri and Herbig~Ae stars from different star-forming regions, the disks in the PRODIGE survey are all T~Tauri systems and located in the same Taurus star-forming region.
Since our disks have many similar properties (e.g., low stellar luminosities, ages, etc.) and have been observed and imaged in the same way, our goal is to present a general model that can be homogeneously applied to the fitting of the selected disks in our sample. This homogenous modeling approach should allow us to better understand the similarities and differences within this class of disks, and how the differences in their structures affect their chemical composition. In Tab.\ref{table:obs_prop} we present an overview of the stellar and disk properties as adopted from \citet{PRODIGE-I}.

\begin{table*}
    \caption{Stellar and disk properties, adapted from \citet{PRODIGE-I}, where further details on the sources can be found.}             
    \label{table:obs_prop}      
    \centering          
    \begin{tabular}{c c c c c c c c c} 
    \hline\hline
    Source & dist. & $M_\star$ & $T_{eff}$ & $\log L_\star/L_\odot$ & PA & incl. & $R_{out}$\\
        &  [pc] & [$M_\odot$] & [K] &  & [deg] & [deg] & [au]\\
    \hline                    
    CI Tau & 159 & 1.0 & 4277 & -0.09 & 282 & 47.3 & 518\\
    CY Tau & 129 & 0.5 & 3560 & -0.61 & 64.5 & 27.1 & 251\\
    DL Tau & 159 & 1.1 & 4277 & -0.19 & 320.3 & 42 & 621\\
    DM Tau & 145 & 0.5 & 3720 & -0.82 & 65.9 & -34.8 & 781\\
    DN Tau & 128 & 0.7 & 3806 & -0.16 & 171.3 & 35.1 & 287\\
    IQ Tau & 131 & 0.6 & 3690 & -0.67 & 311.6 & 60.6 & 212\\
    \hline                  
    \end{tabular}
\end{table*}

% The adopted model is introduced in Sec.\ref{sec:model}, while we now give an overview of the CO isotopologues data introduced by ???Semenov et al. (submitted).

% \subsection{DM Tau}

% \subsection*{DN Tau}
% The stellar mass of the system is $0.87 M_\odot$ and 240~au size in the gas emission and an inner radius of 11~au \citep{Ricci10}. The disk has an inclination of $35.2^\circ$ and PA $185.8^\circ$. The disk mass based on dust emission is $0.01 - 0.04 \; M_\odot$ \citep{Guilloteau16, Ribas20}. ???Semenov et al. (submitted) show a compact CO emitting region ($200-300$~au), with a weak C$^{18}$O~(2-1) emission. The dust emission shows a $\sim 23$~au wide gap at $\sim 60$~au \citep{Long18}. ???Semenov et al. (submitted) find a compact CO emitting region, extending up to 240~au. The peak brightness temperatures are $\sim 9.5$~K for $^{12}$CO~(2-1) and $\sim 1.2$~K for $^{13}$CO~(2-1). The C$^{18}$O~(2-1) is very weak, at $\sim 0.2$~K.

\section{Disk model}
\label{sec:model}
To model the disk structure and chemistry, we developed the \texttt{DiskCheF} framework, of which this work is the first application \citep{Smirnov-Pinchukov22}. This tool was developed to simulate molecular line data using a modular disk physical-chemical framework and can be applied either to fit disk interferometric data or as a standalone disk forward modeling tool. The code philosophy is to establish a series of customizable building blocks so that each code block can be modified without affecting the functionality of the other code structures. A resulting disk model can be built by choosing any combination of the individual disk's physical, chemical, and radiative transfer models. The preset models can be easily subclassed to add custom functionalities when needed. The \texttt{*Base} objects described below are templates showing the information needed by each step of a \texttt{DiskCheF} modeling (namely, the disk physical model, chemical model, radiative transfer model, and fitter). These objects cannot be directly used to create a disk model, so the user has to either use them as a template for the desired theoretical disk model or choose the corresponding default models already implemented in \texttt{DiskCheF}.

The starting point of \texttt{DiskCheF} is to create a disk physical model (\texttt{physics.PhysicsBase}), with information about the gas and dust density and temperature distributions. This model is then used as input for the chemistry model, (\texttt{chemistry.ChemistryBase}), which computes the abundance of the chemical species using the disk physical structure provided by the previous disk physics module. \texttt{DiskCheF} also uses the radiative transfer modeling tool \texttt{maps.RadMCBase} to produce channel and moment maps (\texttt{maps.RadMCRT}) and dust continuum emission maps (\texttt{maps.RadMCRTImage}), and to calculate the disk temperature (\texttt{maps.RadMCTherm}). The workflow of the \texttt{DiskCheF} modeling is shown in Fig.\ref{fig:workflow}. In the next sections, we describe in more detail the individual models implemented within the \texttt{DiskCheF} framework to fit the CO data.

\subsection{Disk physical model}
\label{sec: physical model}
To model the disk's physical and thermal structure, we adopted the \citet{LyndenBell74} prescription. This model is implemented in the \texttt{physics.WB100auWithSmoothInnerGap} module. In this model, the radial surface density distribution of the gas is given by the Lynden-Bell \& Pringle self-similar profile with a smooth inner gap:

\begin{equation}
    \label{eq:LBP}
    \Sigma(r) = \Sigma_0 \, \left(\frac{r}{r_c}\right)^{-\gamma} \, \exp \left[ -\left(\frac{r}{r_c}\right)^{2 - \gamma} \right] \, \exp \left[ -\left(\frac{r}{r_{in}}\right)^{\gamma - 2} \right],\\
\end{equation}

with

\begin{equation}
    \label{eq:LBP_norm}
    \Sigma_0 = \left( 2 - \gamma \right) \frac{M_{disk}}{2 \pi \, r_c^2} \exp \left( \frac{r_{in}}{r_c} \right)^{2 - \gamma},
\end{equation}

\begin{figure*}
\centering
   \includegraphics[scale=0.95]{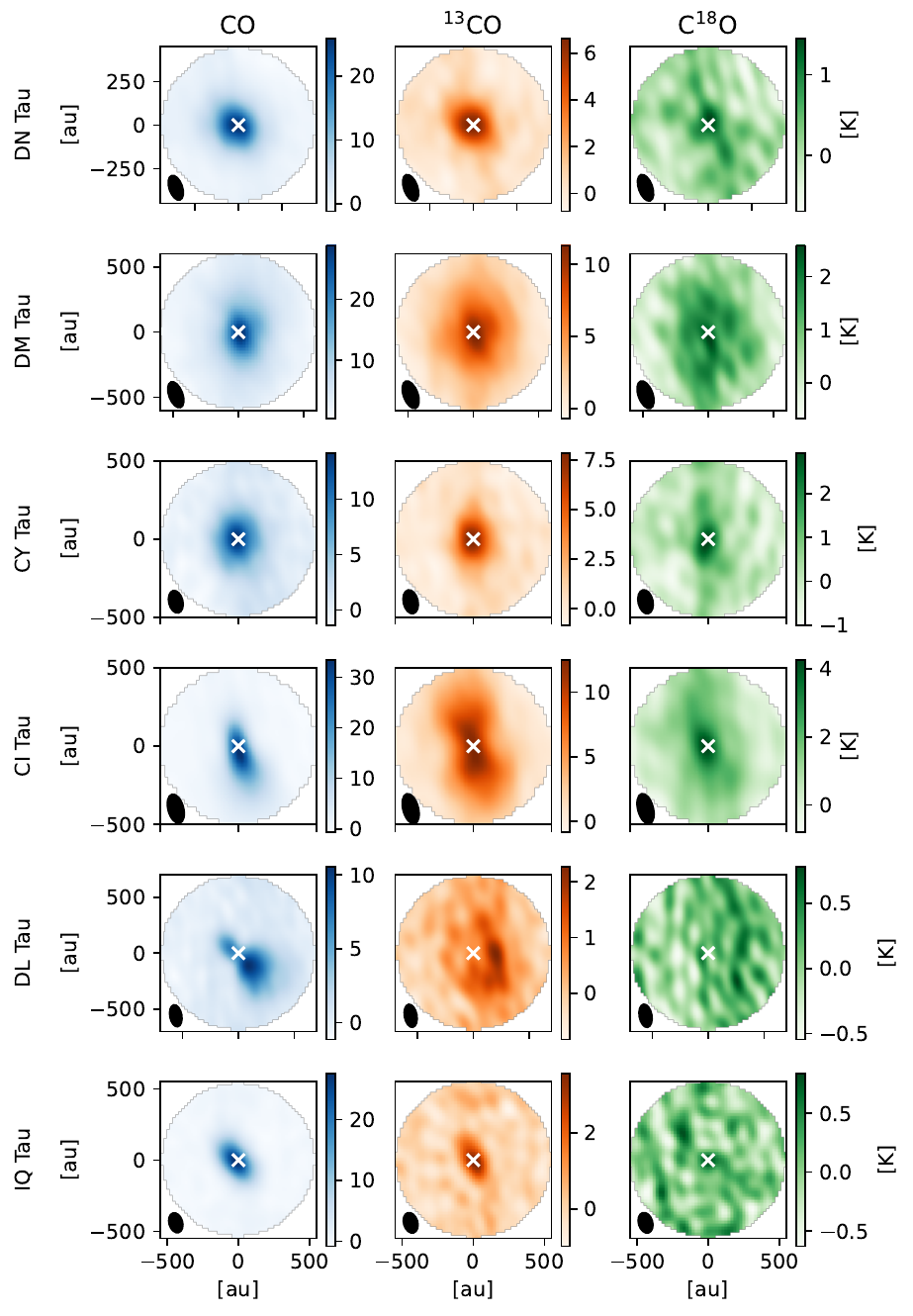}
     \caption{Moment zero maps of the CO, $^{13}$CO and C$^{18}$O~(2-1) emission for the disks chosen for the DiskCheF fitting.}
     \label{fig: mom0}
\end{figure*}

where $\gamma$ is the tapering factor, $r_c$ the tapering radius, $r_{in}$ the inner disk radius, and $M_{disk}$ the total disk mass. The vertical profile is given by integrating the vertical hydrostatic equilibrium equation:

\begin{equation}
    \label{eq:vertical_hydrostatic}
    \frac{\partial \ln \rho}{\partial z} = 
    - \left[ 
        \left(
            \frac{G \, M_{star} \, z}{\left( r^2 + z^2 \right)^{3/2}}
        \right)
        \left( \frac{\mu \, m_H}{k T} \right)
        + \frac{\partial \ln T}{\partial z}
    \right]
.\end{equation}

The midplane and atmosphere temperature of a disk is given by the radial power law distribution:

\begin{equation}
    \label{eq:tmid/atm}
    T_{mid/atm} = T_{mid/atm, \, 100} \,\left(  \frac{r}{100 \, \mathrm{au}} \right)^{-q_{mid/atm}}
,\end{equation}

with $T_{mid/atm, \, 100}$ being the temperature at a radius of 100~au and $q_{mid/atm}$ the exponent of the distribution. The value of the temperature exponent has been measured from the CO data for disks in the Taurus region, and it is usually assumed that a slope of 0.55 for both the midplane and atmosphere temperature profile provides a good representation of the outer disk temperature distribution \citep{Williams14}. The transition between the atmosphere and midplane temperature is parameterized with a sine function:

\begin{equation}
    \label{eq:temperature}
    T_{WB}(r,z) = 
    \begin{cases}
        T_{mid} + \left( T_{atm} - T_{mid} \right) \left[ \sin{\left( \frac{\pi \, z}{2 \, z_q} \right)}\right]^{2\delta}  \hfill \mathrm{if} \; z < z_q,\\
        T_{atm} \hfill \mathrm{if} \; z \geq z_q\,\
    \end{cases}
\end{equation}

where $\delta$ describes the steepness of the profile and $z_q$ is the height over the disk atmosphere where the disk reaches the atmospheric temperature. \citet{Williams14} explored the impact of these parameters and found that these parameters do not significantly affect the line luminosity. In the literature it is usually assumed that $\delta = 2$ and $z_q = 4 \, H_p$, where $H_p$ is the pressure scale height:

\begin{equation}
    \label{eq:hp}
    H_p = \sqrt{\frac{\kappa \, T_{mid} \, r^3}{G \, M_{star} \, \mu \, m_H}},
\end{equation}

where $\kappa$ is the Boltzmann constant, $G$ the gravitational constant, $\mu = 2.3$ the mean molecular weight of the gas, and $m_H$ the mass of atomic hydrogen. Moreover, the typical interstellar radiation field in star-forming regions prevents the disk from reaching unrealistically low temperatures. Following the prescription in \citet{Tazzari21}, we set a threshold of $T_{floor} = 7$~K, and we used an effective temperature of

\begin{equation}
    \label{eq: teff}
    T^4 = T_{WB}^4 + T_{floor}^4.
\end{equation}

\subsection{Chemical model}
The main goal of \texttt{DiskCheF} is to fit disk models to the multi-line and/or multi-species interferometric data. When the data are gas emission lines, we need a chemical model that is fast enough yet feasible enough to be part of a fitting routine. A typical simplified network for the chemical evolution of a protoplanetary disk runs for a few seconds on a single CPU. However, we discussed in Sec.\ref{sec:intro} that we have to account for all of the processes affecting CO abundances, such as self-shielding or photodissociation. The neglect of these processes can underestimate the disk mass by one or two orders of magnitude. Therefore, to fit the CO line emission, we need a more robust chemical network. The computational time for such networks ranges from tens of minutes to days per disk model. Since our fitting routine computes on the order of $10^5$ disk models to converge to a best-fit model, it adds up to unfeasible computational times.

One solution is to apply machine learning techniques to the results of time-dependent chemical kinetics models. These techniques find the correlation between the input and the output of the networks and the smallest set of disk parameters needed to predict chemical abundances without running the network for each disk model.Our chemical predictions are based on the results of the ANDES astrochemical model of a 2D azimuthally symmetric disk \citep{Semenov11}, applying the ALCHEMIC chemical code \citep{Semenov10}. This analysis was performed using the Python machine learning library \texttt{Scikit-learn} \citep{Pedregosa11}. We employed the \texttt{KNeighborsRegressor} estimator, which identifies the k nearest data points in the input feature space and performs the interpolation of the output feature values among them. The full derivation of the correlation between the disk physical parameters and the ANDES CO abundances, as well as the input parameters of the regressor, are described in detail in \citet{Smirnov-Pinchukov22}. Isotopologue selective processes are not included in the chemical network; therefore, we inferred the isotopologue abundances using the standard ISM abundance ratios, CO/$^{13}$CO=69 and C$^{18}$O=557 \citep{Wilson99}.

The full gas-grain chemical network includes about 650 species and 7\,000 reactions, including gas-phase and surface two-body reactions, adsorption and desorption, photoreactions and ionization/dissociation by X-ray, cosmic rays, short-lived radioactive nuclides, and reactive desorption. Following \citet{Eistrup16}, we adopted an icy molecular initial composition based on the abundance of ice in prestellar cores \citep{Oberg11}. We ran the time-dependent chemical evolution til the age of 1~Myr. The disk physical structure was set through a stellar mass, $M_\star$, a disk mass, $M_\mathrm{disk}$, and a disk tapering radius, $r_c$. These parameters define the distribution of densities, temperatures, and the high-energy radiation field. The stellar mass also governs the stellar temperature and luminosity, which were calculated for the age of 1~Myr using the evolutionary model by \citet{Yorke08}. The ionizing radiation field was computed using the \citet{Bruderer09} X-ray prescription and the \citet{Padovani19} cosmic ray prescription. The midplane temperature was computed by integrating stellar and accretion luminosities using a parametric approach, while the atmospheric temperature was computed by solving the vertical UV radiation transfer equation. These two temperatures were then connected using Eq.\ref{eq:temperature}, as in \citep{Williams14}. Iterative calculations of density and temperature distributions were then conducted to achieve a self-consistent solution, as has been outlined by \citep{Molyarova17}.

We created a population of synthetic disks with different $M_\star$, $M_\mathrm{disk}$, $r_c$, and stellar X-ray luminosity, $L_X$, to cover a wide range of physical conditions typical for protoplanetary disks. Relying on the observational constraints on disk structure, these parameters were varied: $M_\star = [0.3, 2.5] \; M_\odot$, $M_\mathrm{disk} = [10^{-3}, 10^{-1}] \times M_\star$, $r_c = [20, 100]$~au, and $L_X = [10^{29}, 10^{31}]$~erg/s. These parameters were used to compute 540 synthetic disk models, where each model is composed of $4\,000$ physical cells. This amounts to $2\,160\,000$ data points, which took about one year of single CPU run time.

\begin{figure*}
\centering
   \includegraphics[width=\textwidth]{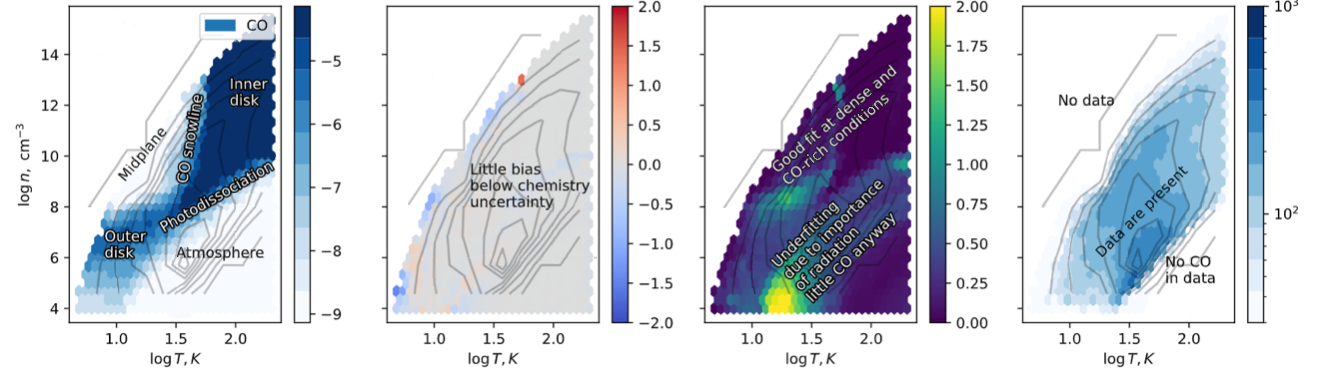}
     \caption{Performance of ML-accelerated chemistry predictions for CO, adapted from \citet{Smirnov-Pinchukov22}. (a): Mean log10 predicted relative abundance as a function of local temperature, gas density, and ionization rate. Darker areas correspond to larger relative (to H atoms) abundances. (b): Median of the difference between the predicted values and test set data (bias, dex), in dex, as a function of temperature and density. Gray areas correspond to an unbiased fit. (c): Standard deviation between the predicted values and test set data, in dex (std, dex). (d): Relative density (histogram) of CO within the data points with contours, also shown in the other panels. Various regions of the protoplanetary disk are described in panel (a).}
     \label{fig: ML prediction}
\end{figure*}

The performance of the machine learning chemical predictions is shown in Fig.\ref{fig: ML prediction} for the gas phase CO abundance, adapted from \citet{Smirnov-Pinchukov22}. The fitting procedure successfully reproduces the chemical model predictions, demonstrating minimal systematic errors in panel (b) and scatter in panel (c). A noticeable increase in scatter is observed within a specific region of the parameter space, corresponding to the radiation-sensitive transition zone between the atmosphere and the other disk regions in the low-density area. This region constitutes only a marginal portion of the total CO inventory, and our conclusions remain unaffected by the scatter in CO abundance predictions within this particular disk region.

The main result of the machine learning analysis of this precomputed disk physical-chemical grid is that the disk CO abundances are mainly determined by the local density and temperature of the gas. Further parameters, such as the ionizing and UV radiation, increase the quality of the predictions but are a minor correction to the abundances predicted by the gas density and temperature alone.  While the impinging FUV radiation does play a role in the chemical evolution, we found that the effect of the local UV field is correlated with the effect of the gas density and temperature. For instance, UV radiation is stronger in the atmosphere, where the gas density is lower and its temperature higher; hence, the CO abundance in this UV-irradiated region can be correctly predicted using the local low gas density and high temperature. In the \texttt{DiskCheF} framework, we used the same prescription for the X-ray and cosmic ray radiation part of the \texttt{physics.PhysicsBase} object and the CO abundances were predicted from the gas density, gas temperature, and ionizing radiation by the \texttt{chemistry.SciKitChemistry} object.

\subsection{Fitting the interferometric line data}
Once we had a physical and chemical disk model, we used RADMC3D to calculate the line radiative transfer of the observed CO isotopologue lines and produce channel maps of these lines with the same spatial and spectral resolution as in the PRODIGE data. However, an interferometer like NOEMA does not produce an image of the source emission, but measures its visibilities, the complex values of its Fourier transform. There are two ways to compare the model channel maps to the data, either by working in the image plane by converting the observed visibilities to an image or by computing the synthetic visibilities of the model by knowing the antenna configuration of the interferometer. While more intuitive, working on the image plane requires several assumptions about the source, and there is no unique way to derive an image from visibility data. This effect is more severe with less sampling of the source visibilities. While NOEMA has a great spectral resolution ideal for deep molecular line surveys, it is more limited in the visibility sampling than ALMA with its $>50$ antennas. 

The second approach is to directly operate in the $uv$-visibility plane, which represents the real instrument data, by taking the spectral resolution, sensitivity, and antenna configuration into account. This provides a more robust comparison with the data, but if used in a fitting algorithm requires the computation of synthetic visibilities at each likelihood evaluation. For this purpose, we used the \texttt{GALARIO} Python library \citep{Tazzari18}, a computationally efficient tool for the generation of synthetic visibilities. In \texttt{DiskCheF}, the object \texttt{uv.UVFits} can be used to read, analyze, and visualize visibility data, including computing the $\chi^2$ of visibility data to a channel map using \texttt{GALARIO} routines (\texttt{uv.UVFits.chi2$\_$with()}). 

With this likelihood evaluation, we could now fit the model to the observations. We adopted the nested sampling Monte Carlo algorithm MLFriends \citep{MLFriends1, MLFriends2}, using the \texttt{UltraNest}\footnote{\url{https://johannesbuchner.github.io/UltraNest/}} package \citep{UltraNest}. This package provides computationally efficient and optimized MPI-cluster tools to find and analyze the posterior probability distribution of the model parameters. This allowed us to find more reliable results than traditional Markov chain Monte Carlo methods with a reasonable amount of computational resources. To fit the PRODIGE data, we deployed 2560 logical CPUs, corresponding to about $5 \times 10^5$ likelihood evaluations within 24 hours of computational time. The disk model that we employed in this first application of \texttt{DiskCheF} has only five free parameters: the total disk mass, $M_\mathrm{disk}$, and the tapering radius, $r_c$, from Eq.\ref{eq:LBP}, and the midplane and atmosphere temperatures, $T_{mid/atm, 100}$, from Eq.\ref{eq:tmid/atm}. The disk model is not that strongly dependent on the other parameters mentioned in Sec.\ref{sec:model}, and we kept them fixed to the values summarized in \citet{PRODIGE-I}.

\section{Results}
\label{sec: results}
\begin{figure*}
\centering
   \includegraphics[width=\textwidth]{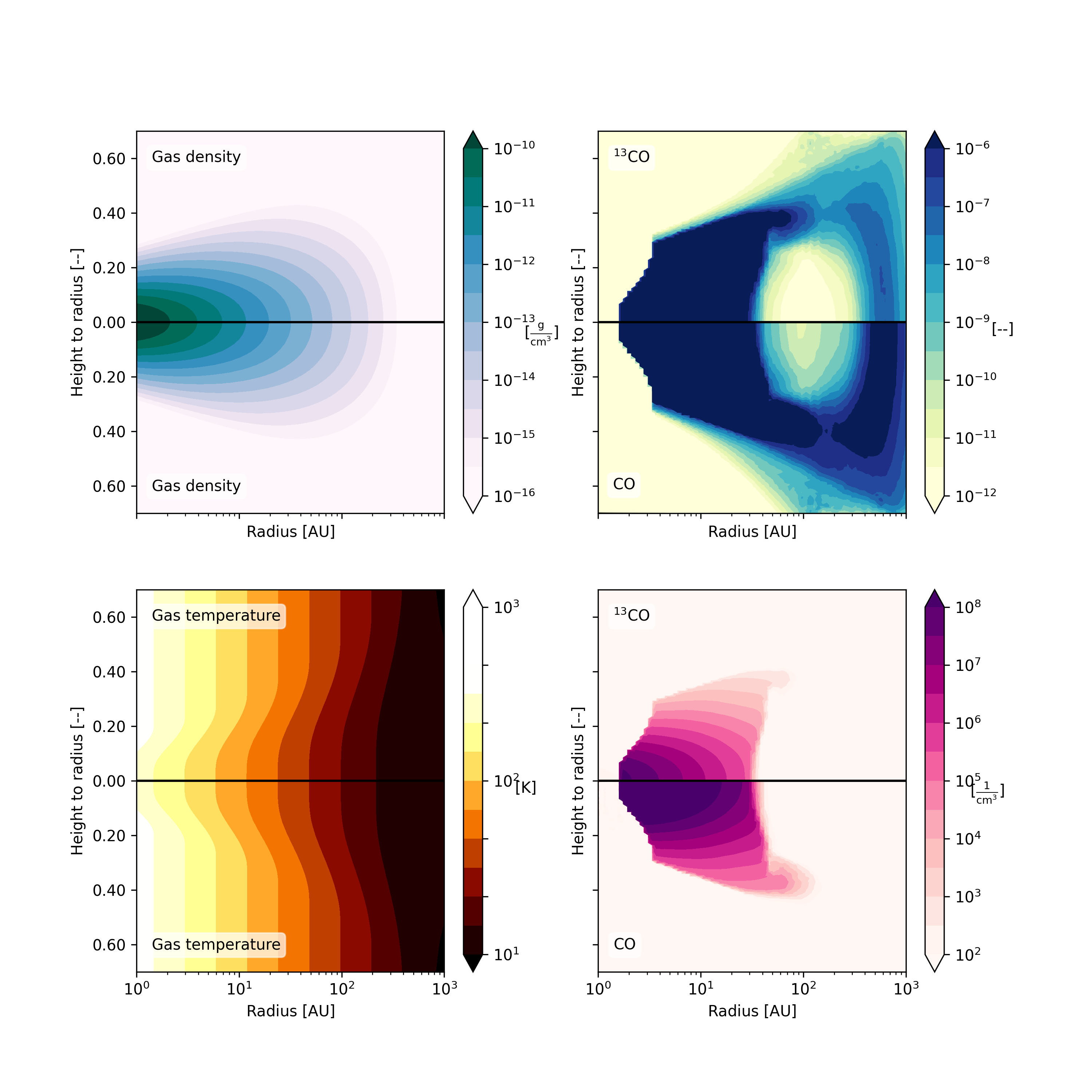}
     \caption{Best-fit physical and chemical model for DM~Tau: (top left) gas volume density distribution, (top right) CO to H$_2$  number density ratio, (bottom left) gas temperature distribution, and (bottom right) CO isotopologues volume density distribution.}
     \label{fig: DM Tau model}
\end{figure*}

\begin{figure*}
\centering
   \includegraphics[width=\textwidth]{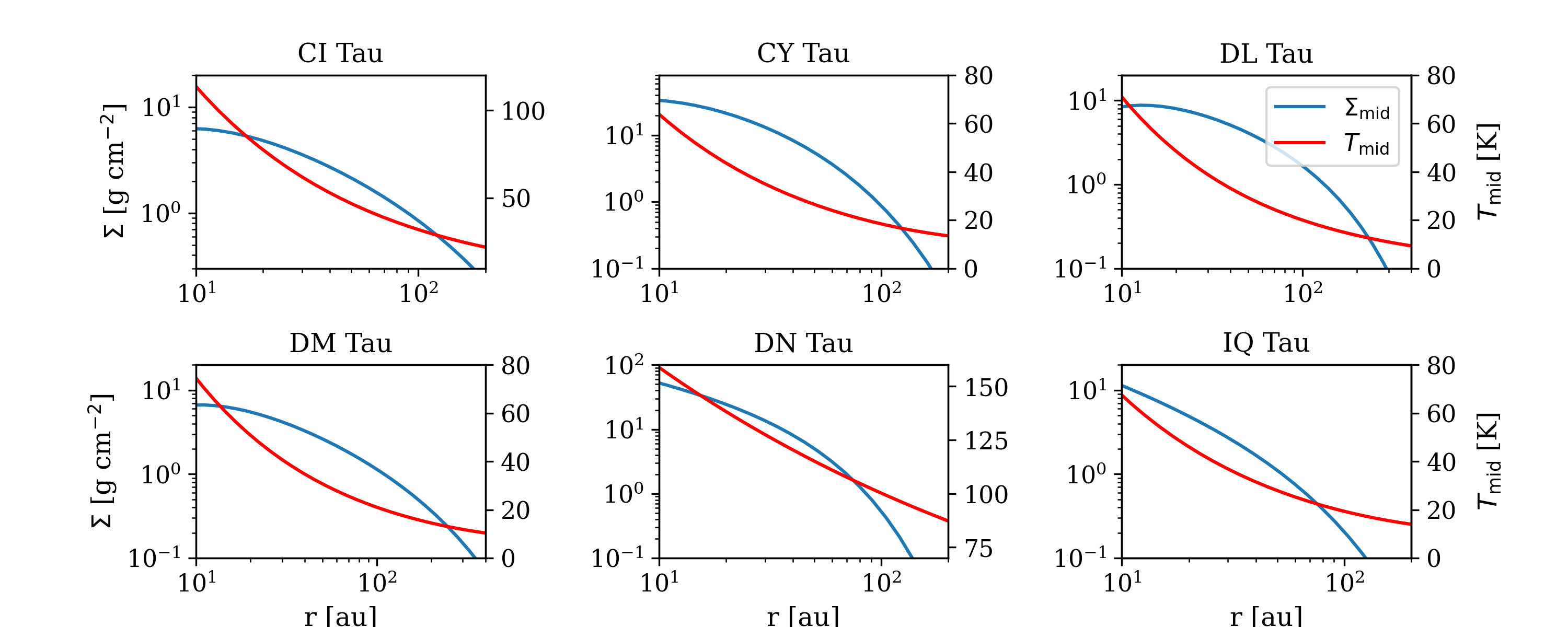}
     \caption{Best-fit column density and midplane temperature profiles for the disks in the PRODIGE sample.}
     \label{fig: profiles}
\end{figure*}

The mass distribution is the most important result of our analysis for planet formation studies. The total disk gas mass can only be traced by optically thin emission. While $^{12}$CO~(2-1) emission is optically thick, and hence can only provide information about the gas temperature at its emission surface, the more optically thin  $^{13}$CO and C$^{18}$O~(2-1) lines can be used to constrain the disk gas mass \citep{DDG03}. To analyze the result of our fitting, we first present the best-fit model for the DM~Tau disk. The data quality of this disk is the highest in our sample since it is the largest and brightest one, and we can use this disk to assess the feasibility and goodness of our best-fit model. In Fig.\ref{fig: DM Tau model} we show the best-fit physical and chemical model for DM~Tau. This figure shows how the ML-based chemical model correctly captures the CO freeze-out from the gas phase at low temperatures in the outer midplane, as well as the CO photodissociation in the disk atmosphere. In Fig.\ref{fig: profiles} the surface density and midplane temperature profiles of our best-fit models are shown.

HD- or CO-based mass estimates can be poorly constrained due to the uncertainties in the underlying gas thermal structure. However, our approach of fitting at the same time optically thin and optically thick CO isotopologue lines allows us to constrain both the disk mass and temperature distribution, leading to more reliable mass estimates. \cite{Law21} found that the temperature profiles of the disks in the MAPS sample follow the power law in Eq.\ref{eq:tmid/atm} outside the 100-150~au radius, but have a flatter profile in the inner disk. Thus, two power law profiles would provide a better fit for the high-resolution ALMA MAPS data.

\begin{figure*}
\centering
   \includegraphics[width=\linewidth]{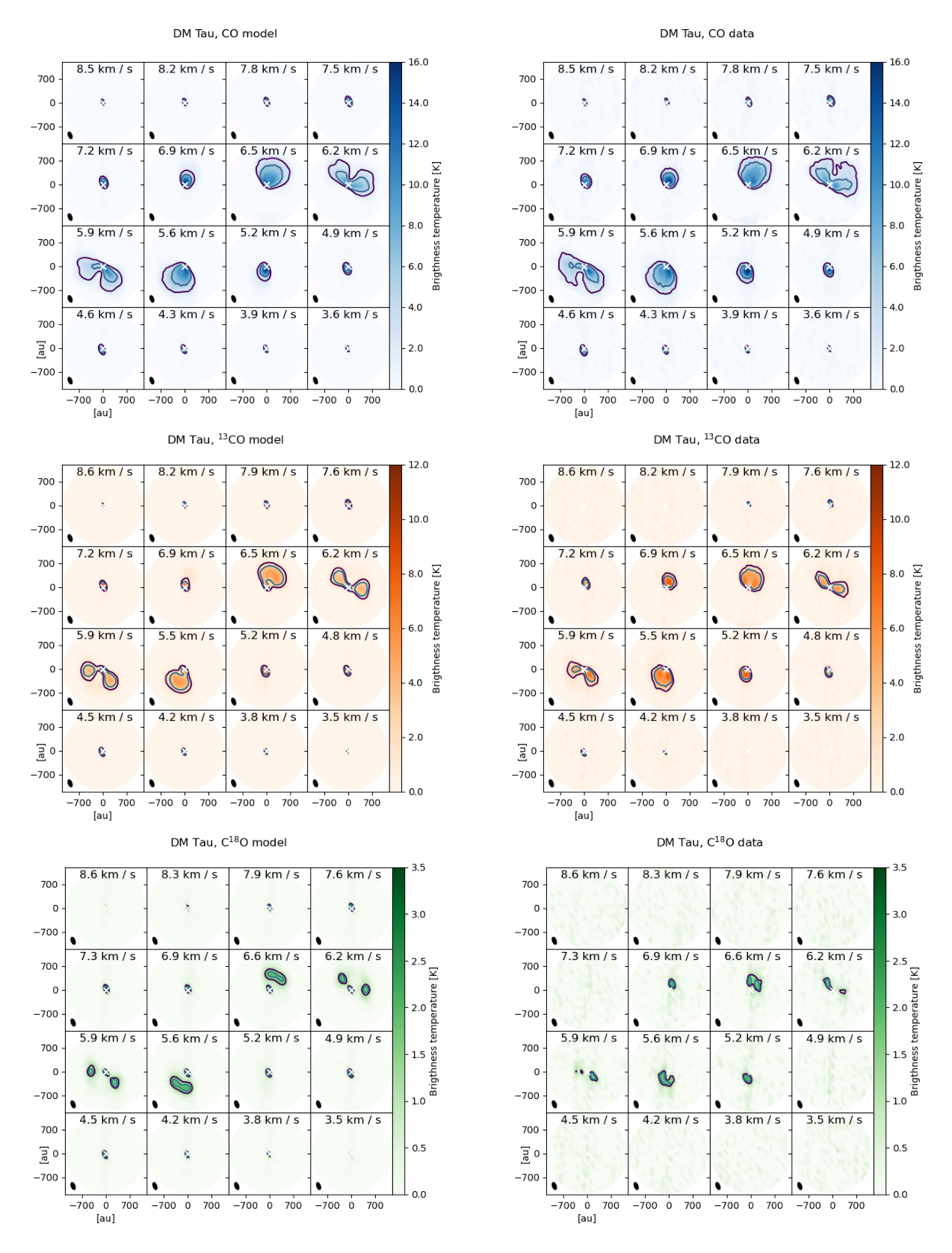}
     \caption{Comparison between the best-fit model and the observations of DM~Tau for the CO, $^{13}$CO, and C$^{18}$O isotopologue emission.}
     \label{fig: DM Tau images}
\end{figure*}

\begin{figure}
\centering
   \includegraphics[height=\textheight]{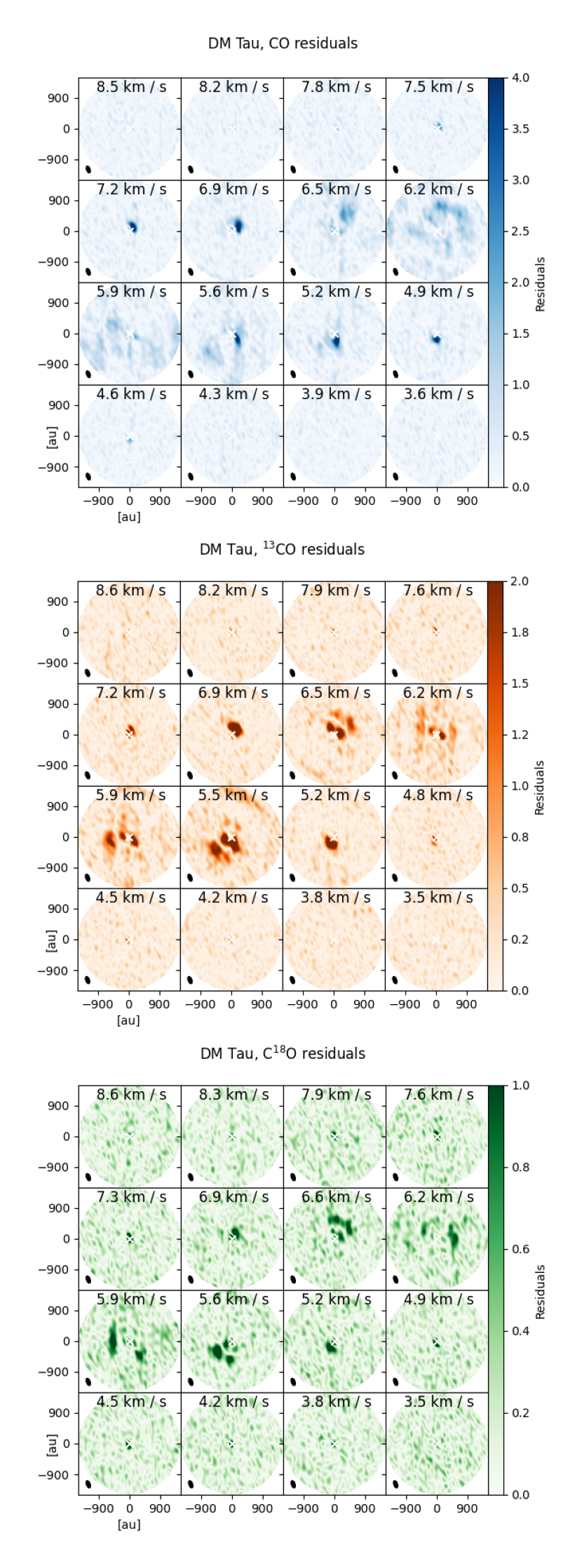}
     \caption{Residuals of the DM~Tau best-fit model for the CO, $^{13}$CO, and C$^{18}$O isotopologue emission.}
     \label{fig: DM Tau residuals}
\end{figure}

However, the spatial resolution in our sample is about the size of the inner disk, and we cannot properly constrain the temperature profile in this region. Therefore, we assume a single power law distribution for the whole disk structure. We demonstrate the effect of this assumption on the DM~Tau disk. In Fig.\ref{fig: DM Tau images} we show the comparison of the best-fit model and the data in the CO~(2-1) isotopologue channel maps, while in Fig.\ref{fig: DM Tau residuals} the model-data residuals are shown. The residuals are larger in the inner disk, where the model underestimates the gas brightness temperature by about 2~K, and where the CO emission becomes optically thick, and hence more sensitive to the gas temperature than the gas density. However, the mass distribution is not affected by this assumption, since it is more strongly constrained by the outer disk emission, and thus we still get a reliable gas mass estimate.

\begin{table*}
    \caption{Best fit of the disk physical parameters, compared with the mass estimates from other methodologies, as is discussed in the text.}
    \label{table: fit results}      
    \centering          
    \begin{tabular}{c c c c c | c c}
    \hline\hline       
    Source  &   $r_c$    &   $T_{atm,\,100}$ &   $T_{mid,\,100}$ & $M_{disk}$ & $M_\mathrm{dust} \times 100^{(1)}$ & $M_{WB}^{(2)}$\\
        &   [au]   &   [K] &   [K] &  [$10^{-2} \times M_\odot$] & [$10^{-2} \times M_\odot$] & [$10^{-2} \times M_\odot$]\\
    \hline
    CI~Tau &  150 & 35 & 32 & 2.0 & 1.6 & 0.3\\
    CY~Tau &  40  & 22 & 18 & 1.8 & 1.9 & 0.1\\
    DL~Tau &  125 & 24 & 20 & 2.5 & 2.5 & 0.005\\
    DM~Tau &  200 & 31 & 21 & 4.0 & 2.5 & 0.09\\
    DN~Tau &  40  & 27 & 20 & 2.0 & 1.3 & - \\
    IQ~Tau &  50  & 27 & 19 & 0.5 & 0.9 & 0.007\\
    \hline
    \end{tabular}
    \tablebib{(1)~\citet{Guilloteau11, Guilloteau16, Guedel18}, (2)~\citet{Williams14}}
\end{table*}

The CO~(2-1) line intensity of the output images is strongly affected by the disk temperature profile. Moreover, changes in the disk mass do affect the shape of the CO~(2-1) emission lines as well, since the higher disk mass changes the Keplerian velocity profile of the CO gas. Indeed, in more massive disks there is more mass at higher heights over the midplane, and the resulting projection of the Keplerian velocity profile is noticeably different. The $^{13}$CO~(2-1) data, on the other hand, are more sensitive to the mass distribution than the gas temperature, as we would expect from this (partly) optically thin emission. Therefore, $^{13}$CO~(2-1) data provide a more reliable disk gas mass estimate when fitted together with the $^{12}$CO~(2-1) emission probing mainly the disk thermal structure. The C$^{18}$O~(2-1) images follow the same trend as the $^{13}$CO images but, being much fainter, this line is not always detected with a good-enough signal-to-noise ratio to provide additional good constraints on the disk gas mass.

%However, an inner gap in the disk structure can affect the mass estimate derived from this density profile. Since the gas is most dense in the inner disk, we find here most of the disk mass. A wrong estimate of the disk's inner radius can therefore lead to a wrong disk mass estimate, even if the shape of the distribution is properly constrained.

Another parameter affecting the results of the \texttt{DiskCHeF} fitting is the depletion of CO. Our predictions are based on the results of the gas-grain ANDES2 chemical model, which does not include isotopologue-selective CO photodissociation processes \citep[e.g.,][]{Visser09}. This may lead to an underestimation of the disk masses probed via the minor CO isotopologue lines, and our estimates could be a lower limit \citep{Miotello14, Miotello16}. This is a feature of the machine learning method adopted to speed up chemical modeling, as the ML-chemistry results inherit the same limitations that the chemical model used to produce the training data. The machine learning-accelerated chemistry based on the ANDES2 model made it possible for us to add a complex gas-grain chemical kinetics model into our fitting algorithm, instead of adopting a parameterized chemical model \citep{Williams14} or a reduced CO network \citep{Deng_ea23}. One of the main results of this paper is the application of the fast machine learning method to accelerate calculations of molecular abundances, which can be easily implemented within the flexible DiskCheF framework, using other, disk chemical kinetics models such as DALI or ProDiMo.

We expect the CO~(2-1) emission to be in the local thermodynamical equilibrium (LTE) with the gas at the disk densities and temperatures from which it is emitted. The best-fit DM~Tau model has CO~(2-1) brightness temperatures matching the thermal profile of the underlying best-fit model, with a peak brightness temperature over the beam-emitting area of 17~K. In the other disks, the peak brightness temperature over the beam area ranges from 11~K in IQ~Tau to 20~K in CI~Tau. On the other hand, the $^{13}$CO and C$^{18}$O~(2-1) emissions are optically thin, with a ratio of $^{13}$CO-to-C$^{18}$O emission equal to the isotopologue ratio (about 7.3). However, in the inner disk of the more massive disks, the emission ratio falls below the isotopologue ratio (about five in the most massive disk, DM~Tau). This is an indication that the $^{13}$CO emission becomes partially optically thick in the inner disk as well (smaller than our beam size, about 100~au). The C$^{18}$O emission remains optically thin at our spatial resolution, and our mass estimate of 0.04~$M_\odot$ is not affected by this opacity effect.

\subsection{Foreground cloud absorption}
The optically thick CO emission observed in CI, CY, DG, and DL~Tau is affected by foreground absorption within the velocity channel range of approximately 4 to 6~km/s. Given that these disks exhibit a local standard of rest (LSR) velocity of around 6 km/s, the foreground cloud impacts the gas signature in the outer disk as it obscures the red-shifted emission from the Keplerian disk. This obstruction results in the asymmetrical features evident in the moment zero maps presented in Fig.\ref{fig: mom0}.

To find the best fit in this case, we excluded these channels from our likelihood estimate. In doing so, we emphasize the inner disk emission in determining the best-fit model. This choice is made with consideration of the discussion in Section \ref{sec: disk vertical structure}, where we acknowledge that the inner and outer disk structures can significantly differ, particularly in cases involving a flared outer disk structure. This discrepancy introduces uncertainty into the mass estimation for these partially obscured disks when using optically thin isotopologues. This is because the temperature in the outer disk is deduced from the inner disk structure, and the mass estimate from optically thin CO isotopologues is temperature-dependent.

\subsection{Disk vertical structure}
\label{sec: disk vertical structure}
The disk physical model used to fit the data, described in Sec.\ref{sec: physical model}, reconstructs the gas vertical distribution assuming hydrostatic equilibrium. However, recent studies have shown how the gas emission layer may differ from the one determined by the hydrostatic equilibrium. \cite{Law21} found that the flaring indices of the CO emission surfaces for the disk in the MAPS sample can be larger than the one coming from hydrostatic equilibrium by a factor of a few. This effect was previously studied in the case of IM~Lup, a particularly large and flared disk \citep{Pinte18, Franceschi23}. Similar results have been found for the largest disk in our sample, DM~Tau \citep{DDG03}. The shape of the emitting region of CO isotopologues is determined by the stellar irradiation, rather than the disk scale height, tracing the surface of optical depth $\tau \sim  \frac{2}{3}-1$ with respect to the stellar irradiation, and this can affect our temperature estimate. Our best-fit model in hydrostatic equilibrium effectively reproduces the observational data for all the disks in our survey. However, a more sophisticated physical model may be necessary to study the disk structure in more detail, such as trying to model other molecules or by including dust continuum emission data into the fitting.

\subsection{Temperature profiles}
\label{sec: temperature profile}
In our initial modeling attempts, we introduced two additional fitting parameters: the exponent governing the midplane and atmospheric temperature distributions, denoted as $q_{mid/atm}$ in Eq.\ref{eq:tmid/atm}. However, our fitting algorithm revealed a complete degeneracy between $T_{mid}$ and $q_{atm}$, as well as between $T_{atm}$ and $q_{mid}$. This can be attributed to the fact that the observed flux arises from the integrated emission along the line of sight, rendering no formal distinction between the emission originating from a cold midplane and a warm atmosphere, or a warmer midplane and a colder atmosphere in the model and, hence, in our channel map data. This degeneracy could potentially be resolved through either higher-angular-resolution data, enabling precise measurements of the height of the emitting surface for each velocity channel, as has been suggested by previous studies (e.g., \citealt{Dullemond20, Law21}), or by adopting a more realistic underlying disk physical model.

We have adopted a commonly employed assumption in the literature and set the temperature exponent, $q_{mid/atm}$, to a fixed value of 0.55. Protoplanetary disks, such as those found in the PRODIGE sample and characterized by the absence of severe substructures or transient accretion outburst phenomena, generally exhibit temperature slopes in the range of 0.4 to 0.6 \citep{Dutrey14, Law21, Zhang21, Miotello22}. Given the relatively limited variation in temperature slope and our moderate angular resolution, minor adjustments to the temperature profile are unlikely to significantly impact our best-fit results.

\subsection{Disk masses}
The emission of CO isotopologues combined with the results of the ALCHEMIC chemical code allows us to derive the total disk masses. The estimates of the disk gas masses are summarized in Tab.\ref{table: fit results}, along with previous dust-based estimates. Generally, the masses obtained from our methodology align with the dust masses derived from continuum emissions, assuming a dust-to-gas ratio of 100, as was measured in \citet{Guilloteau11, Guilloteau16, Guedel18}. Notably, exceptions are observed in the cases of DM~Tau and DN~Tau, where our mass estimates are about twice the values derived from dust emissions.

Several disks in our sample have mass estimates derived from alternative methodologies. For instance, \citet{McClure16} computed a mass range of $[1.0-4.7]\times10^{-2}$M$_\odot$ for the DM~Tau disk based on HD emission, in accord with our estimate of $4.0\times10^{-2}$M$_\odot$. Another independent measurement for the CY~Tau disk mass was provided by \citet{Powell19}, which relates the size of the large grain-emitting region to the total gas mass of the disk. Their estimation of a substantial disk mass, 0.1~M$_\odot$, approximately 20\% of the host star mass, contrasts with our estimate, which is approximately five times lower. This discrepancy can be explained by \citet{Franceschi22}: the method used by \citet{Powell19} tends to overestimate the disk mass by up to one order of magnitude, justifying the divergence from our results.

The disks in our sample (except for DN~Tau) have been previously characterized for their CO isotopologues emissions using a parametric model for the CO abundance by \citet{Williams14}. These mass estimates are also reported in Tab.\ref{table: fit results}. These values are approximately one order of magnitude lower than our measurements, indicating a gas-to-dust ratio close to ten. One plausible explanation for this disparity is the assumption made by \citet{Williams14} that the CO abundance in the warm molecular layer is equivalent to that in the ISM ($x_\mathrm{CO} = 10^{-4}$). However, processes such as the conversion of CO to CO$_2$, ice chemistry driven by ultraviolet (UV) radiation, dust evolution and vertical transport, and isotope-selective photodissociation can lower the gas-phase CO abundances in disks considerably \citep[e.g.,][]{Trapman_ea21a, Zhang_ea21_MAPS, Furuya_ea22a, TvS_ea22, VanClepper_ea22}, leading to a higher total mass estimate from the same observed CO isotopologue emissions.

In the case of the DM~Tau disk, \citet{Williams14} infer a higher gas-to-dust ratio of 60 and a disk gas mass of $0.9 \times 10^{-2}$~M$_\odot$, aligning their measurement more closely (approximately 25\%) with our estimate and consistent with the lower edge of the mass range inferred from HD emission \citep{McClure16}. This observation suggests that CO destruction processes are less pronounced in the bulk of the DM~Tau structure. Indeed, our model indicates that this disk is warm, reducing the amount of CO depleted from the gas phase by freeze-out, and massive, increasing the amount of CO molecules surviving photodissociation through self-shielding. These effects converge to bring the CO abundance closer to the ISM value, thereby mitigating the disparity between our results and the mass estimate by \citet{Williams14}. 

Indeed, \citep{Trapman22} used thermochemical models to contain the CO abundance in the gas phase in the DM~Tau, finding that $x_\mathrm{CO} = [0.23-1.3] \times 10^{-4}$, supporting our argument that CO abundance in the gas phase is not much lower than its ISM value. Using this $x_\mathrm{CO}$ estimate, \citet{Trapman22} find that the total mass of the DM~Tau falls between $[3.1-9.6] \times 10^{-2}$~M$_\odot$, in agreement with our estimate of $4.0\times10^{-2}$M$_\odot$. Our results for DM~Tau then align with both HD-based mass measurements and CO depletion studies based on thermochemical models, proving the robustness of our methodology for this particular source. This suggests that our findings are likely reliable for other sources as well, although additional independent mass measurements for the remaining sources in our sample would further demonstrate the robustness of our method.

\section{Discussion}
\label{sec:diss}
We employed our \texttt{DiskCheF} fitting model to analyze the disks within our sample, systematically evaluating the fidelity of our best-fit models in replicating the observed data. A comprehensive model-data comparison for all sources is presented in the appendix for brevity, while an overview of the derived best-fit disk physical parameters is presented in Table \ref{table: fit results}. It is worth noting that the mass estimates of our best-fit models are contingent upon both the luminosity of the CO emission and its kinematic properties, which collectively define the morphology of the emission across each velocity channel. The fitting of individual velocity channels consequently provides robust constraints on the disk masses.

A limitation shared by all our best-fit models pertains to the brightness of the CO isotopologues within the initial few tens of~au in the inner disk. This luminosity is consistently brighter than what is observed in the data. This discrepancy can be attributed to the assumed power-law prescription for the temperature distribution. As is discussed in Sec.\ref{sec: temperature profile}, this particular temperature distribution tends to overestimate temperatures within the inner disk, thereby leading to elevated residuals in this region. Nevertheless, most of the disk mass is found in the outer disk, and this discrepancy does not significantly affect our mass estimation.

Our estimates of disk mass align well with previously reported values found in the literature, summarized in Tab.\ref{table:obs_prop}, which are derived from dust emission analyses. An outlier is the DM~Tau disk, which is the most massive source in the sample and much more radially extended even when compared to the DL~Tau disk, which shares a similar mass. Disks with these characteristics are subjected to more rapid dust evolution processes. Consequently, they tend to exhibit higher dust-to-gas ratios than the commonly assumed value of 100, which is employed to estimate disk masses in Table \ref{table:obs_prop} (e.g., \citealt{Franceschi23}). This heightened dust-to-gas ratio justifies why our model mass estimate is double the value of the dust-based mass estimate from Tab.\ref{table:obs_prop}.

Our best-fit models effectively reproduce the observed emission within each channel map, reproducing both the CO line brightness and spatial distribution. The CI~Tau, CY~Tau, and DL~Tau disks are, however, a notable exception. In these particular sources, our models reproduce well the observed $^{12}$CO~(2-1) emission, not the observed optically thin minor CO isotopologue emission lines. This discrepancy does not originate from a wrong disk mass estimate, as modifying the mass would change the kinematic gas distribution, which, in contrast, is well matched by our models. This deviation is also not a consequence of temperature-related effects. Raising the atmospheric temperature, for instance, would result in an overly intense CO emission, while increasing the midplane temperature would lead to a similar outcome for the C$^{18}$O emission. The only plausible explanation for this behavior, where optimizing the fit for one isotopologue negatively impacts the fit for another, is the assumption of an incorrect isotopologue ratio.

Our fitting routine, based on assumed CO isotopologue ratios for these sources, boosts the $^{13}$CO line by increasing its temperature, and since this isotopologue is more optically thin than CO, it is more sensitive to the midplane temperature, which is increased to values similar to the disk atmospheric temperature. The disk atmosphere is usually warmer by 10-20~K in the outer, $\gtrsim 100-200$~au T~Tau disks, and the anomaly in the temperature profile can be used as a signature of a wrongly assumed $^{12}$CO/$^{13}$CO isotopologue ratio. Indeed, in the instance of the CI~Tau disk, we refined our model fit to the observational data by increasing the $^{13}$CO isotopologue ratio by a factor of five compared to the $^{13}$C/$^{12}$C ratio. Simultaneously, we adjusted the C$^{18}$O abundance, reducing it by a factor of three in relation to the $^{18}$O/$^{16}$O ratio. In the case of the CY~Tau and DL~Tau disks, we find that optimizing the $^{13}$CO isotopologue ratio by doubling it and adjusting the C$^{18}$O isotopologue ratio by halving it yields a superior alignment with the data. 

It is crucial to highlight that our approach does not rely solely on the isotopologue brightness profile to constrain the disk mass. Instead, it also considers their spatial distribution across each velocity channel, determined by the gas kinematic structure. The gas kinematics depend upon the disk mass, influencing both the brightness of the optically thin isotopologues and their spatial distribution.

According to our model, a correct prediction of the gas kinematic structure with an inaccurate optically thin isotopologue emission with respect to the observations is more plausible than a model with the wrong kinematic structure and accurate optically thin isotopologue emission. Therefore, the isotopologue ratios primarily impact the gas temperature estimate rather than the already constrained disk mass determined by gas kinematics. Consequently, adjustments to the isotopologue abundances do not significantly impact our overall mass estimates.

We stress that these estimates of the CO isotopologue ratios do not stem from our fitting algorithm, as the depletion factor is a degenerate parameter with the gas density and temperature. To obtain more feasible fits to the CO disk data with \texttt{DiskCheF} model, it is imperative to train the ML algorithm using a chemical network that accounts for $^{13}$C and $^{18}$O-fractionation and isotope-selective photodissociation, following the methodology described in this work. 

It is interesting to note that previous modeling studies focused on fitting the observed CO emission in disks have also inferred lower $^{12}$CO/$^{13}$CO ratios of $\sim 20-40$ than the local ISM $^{12}$C/$^{13}$C isotopic ratio of about 69 \citep[for a recent comprehensive review, see][]{Nomura_ea23}.
Using the Plateau de Bure multi-line, CO isotopologue observations, \citet{Pietu_ea07} have derived the $^{12}$CO/$^{13}$CO ratio of $\sim 20$ in the cold outer DM~Tau T~Tauri disk. Recently, \citet{Yoshida_ea22} have also derived a similar $^{12}$CO/$^{13}$CO ratio of $\sim 20$ in the outer T~Tauri disk around TW~Hya, while in the inner, $\sim 20$~au region of the TW~Hya disk this $^{12}$CO/$^{13}$CO ratio seems to be higher, $\sim 40$ \citep{Zhang_ea17}.
Furthermore, the H$^{13}$CO$^+$/HC$^{18}$O$^+$ ratio of $\sim 8$ has been measured in the TW~Hya disk by \citet{Furuya_ea22b}, which coincides with the local ISM value within the uncertainties. 

The reason why the $^{12}$CO/$^{13}$CO ratio measured in the T~Tauri disks could be lower than the local ISM value is carbon fractionation in the cold, $\sim 20-25$~K gas, typical for the outer regions of these disks probed with CO rotational lines \citep[e.g.,][]{Woods_Willacy09,Nomura_ea23}. The key fractionation reaction is $^{13}$C$^+$ $+$ $^{12}$CO  $\rightarrow$ $^{13}$CO + $^{12}$C$^+$ $+$ $\Delta$E, where ionized atomic carbon is produced by the cosmic ray ionization and the energy barrier, $\Delta$E, is about 35~K. The forward pathway of this reaction is more efficient than the backward route in the gas with $T \lesssim 20-30$~K, which corresponds to the outer, $\gtrsim 50-100$~au regions of T~Tauri disks (the densest, bottom part of the outer, $r \gtrsim 100$~au molecular layer). Indeed, in the theoretical modeling of \citet{Woods_Willacy08}, it has been found that the $^{12}$CO/$^{13}$CO ratio could become as low as about 25 in a T~Tauri disk due to this process. More detailed studies of carbon fractionation in a statistically significant sample of disks are needed to further understand the efficiency of this fractionation and the magnitude of its impact on the observed $^{12}$CO/$^{13}$CO ratios and, hence, disk gas mass estimates using the CO disk emission.

\section{Conclusions}
The gas mass of a protoplanetary disk is an important but challenging quantity to measure. The CO emission remains the most reliable gas tracer, but CO-based mass estimates rely on the assumed gas temperature distribution and disk chemistry. A combination of optically thin and optically thick emission lines, however, can solve the degeneracy between the temperature and the abundance of CO isotopologues emission, with the help of chemical models. In this work, we propose a machine learning-based chemical model to predict a disk chemical composition from its physical structure without the need to run a chemical network. This makes it possible to reproduce observations of line emission through theoretical models using a fitting algorithm, which would otherwise require unreasonable computational resources. We tested the applicability of this method using the CO isotopologue emission data from the PRODIGE program on the NOEMA instrument. This data consists of CO, $^{13}$CO, and C$^{18}$O isotopologue emission from six Class~II protoplanetary disks from the Taurus region. Our findings can be summarized as:

\begin{itemize}
    \item The combination of optically thin and optically thick CO isotopologue emission, combined with the kinematic gas distribution in each velocity channel, can properly constrain the disk physical structure and is in good agreement with previous dust-based mass estimates. This approach gives tighter constraints on the disk's thermal structure and provides a more reliable mass estimate.
    \item There is evidence that the emission surface of optically thick lines does not trace the gas pressure scale height, especially for a large disk with evidence of a flared structure, such as DM~Tau. The vertical gas structure could not be in hydrostatic equilibrium in the inner disk, making it challenging to reproduce the inner and outer disk emission with a single temperature profile. However, most of the disk mass is found in the outer disk, and this effect does not significantly affect our mass estimates.
    \item The cloud absorption obscures the optically thick emission in the channels close to the rest frequency for a few disks in our sample. These channels show the emission coming from the outer regions of the disk and give us information on the temperature structure. For these obscured disks, the temperature structure is constrained only by the inner disk emission, which could have a different profile than the outer disk. In a particularly flared structure, the difference between the inner and outer disk structures adds to the uncertainty of the mass estimate.
    \item In some sources, the CO isotopologue abundances are determined by isotopologue-selective processes. In these sources, the mass is still well constrained by the gas kinematic structure, but the temperature profile is offset to boost the emission of under-abundant isotopologues, and we get a worse fit to the data. By changing the isotopologues ratios, we get a better fit to the data, allowing us to estimate the isotopologue abundances in these sources.
\end{itemize}

\begin{acknowledgements}
This work is based on observations carried out under project numbers L19ME and S19AW (for DNTau and CI Tau) with the IRAM NOEMA Interferometer. IRAM is supported by INSU/CNRS (France), MPG (Germany), and IGN (Spain). D. S., Th. H., G. S.-P., R. F, K. S., and S. vT acknowledge support from the European Research Council under the Horizon 2020 Framework Program via the ERC Advanced Grant No. 832428-Origins. This work was partly supported by the Programme National “Physique et Chimie du Milieu Interstellaire” (PCMI) of CNRS/INSU with INC/INP co-funded by CEA and CNES. This research made use of NASA’s Astrophysics Data System. 
\end{acknowledgements}


\bibliographystyle{aa.bst} % style aa.bst
\bibliography{aanda.bib} % your references Yourfile.bib
\onecolumn
\begin{appendix}
\section{Model-data comparison and residual channel map}
\label{appendix}
\begin{figure*}[h]
\centering
   \includegraphics[height=0.9\textheight]{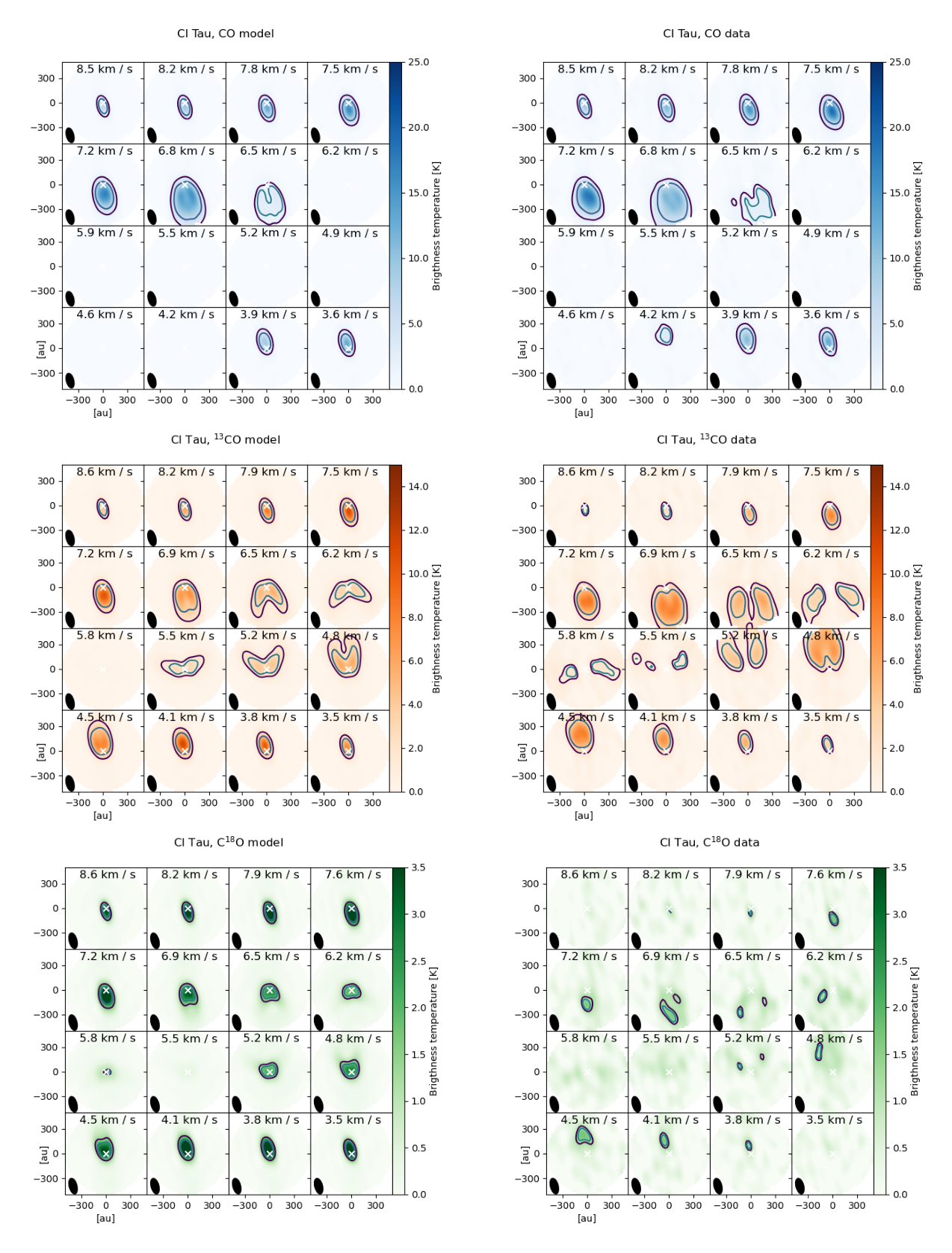}
     \caption{Comparison between the best-fit model and the observation of CI~Tau for the CO, $^{13}$CO, and C$^{18}$O isotopologue emission.}
     \label{fig: CI Tau images}
\end{figure*}

\begin{figure*}
\centering
   \includegraphics[height=0.9\textheight]{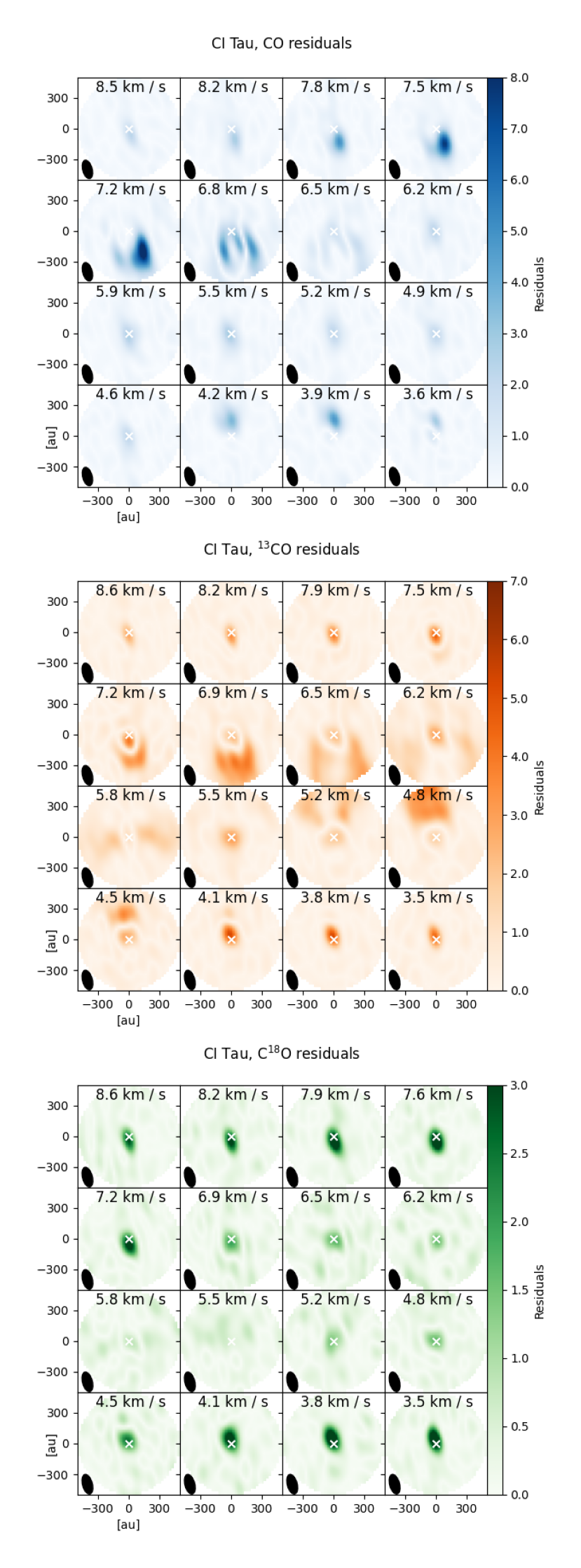}
     \caption{Residuals of CI~Tau best-fit model for the CO, $^{13}$CO, and C$^{18}$O isotopologue emission.}
     \label{fig: CI Tau images residuals}
\end{figure*}

\begin{figure*}
\centering
   \includegraphics[height=0.9\textheight]{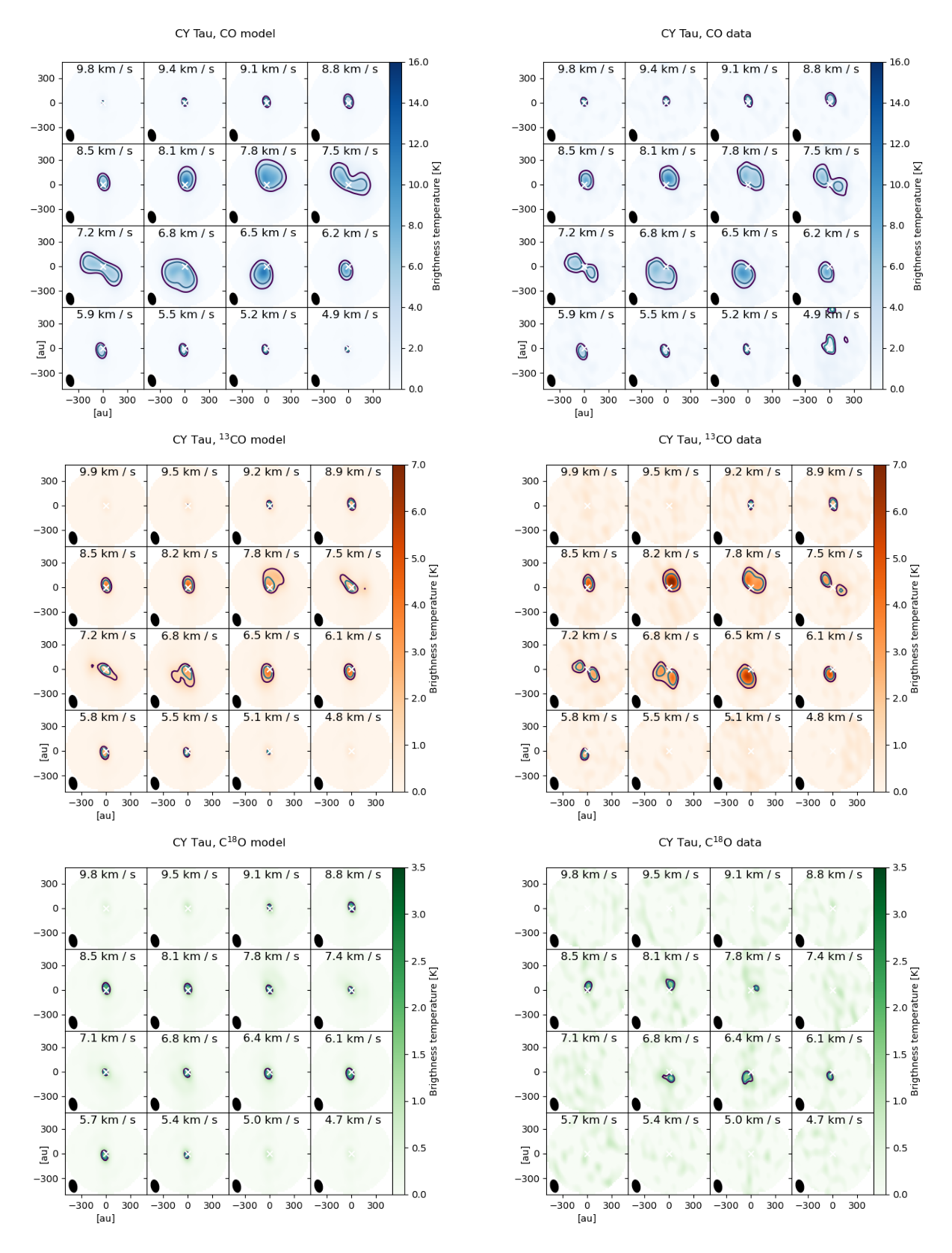}
     \caption{Comparison between the best-fit model and the observation of CY~Tau for the CO, $^{13}$CO, and C$^{18}$O isotopologue emission.}
     \label{fig: CY Tau images}
\end{figure*}

\begin{figure*}
\centering
   \includegraphics[height=0.9\textheight]{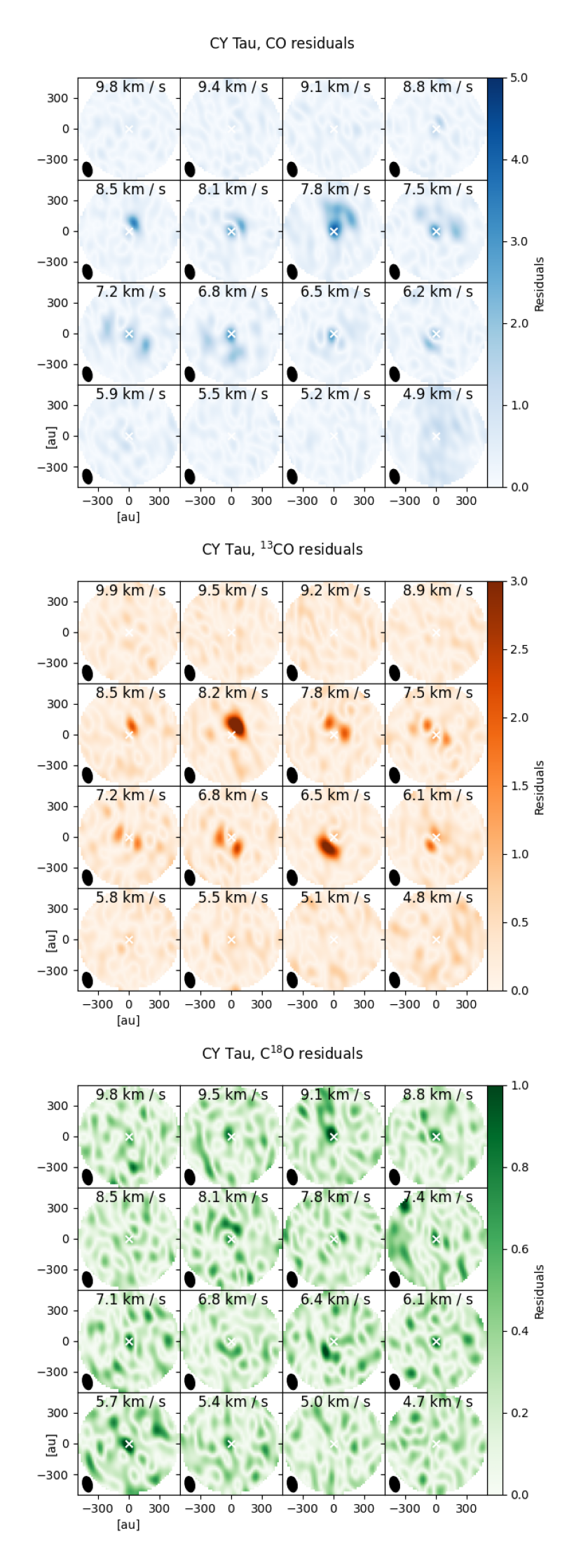}
     \caption{Residuals of CY~Tau best-fit model for the CO, $^{13}$CO, and C$^{18}$O isotopologue emission.}
     \label{fig: CY Tau images residuals}
\end{figure*}

\begin{figure*}
\centering
   \includegraphics[height=0.9\textheight]{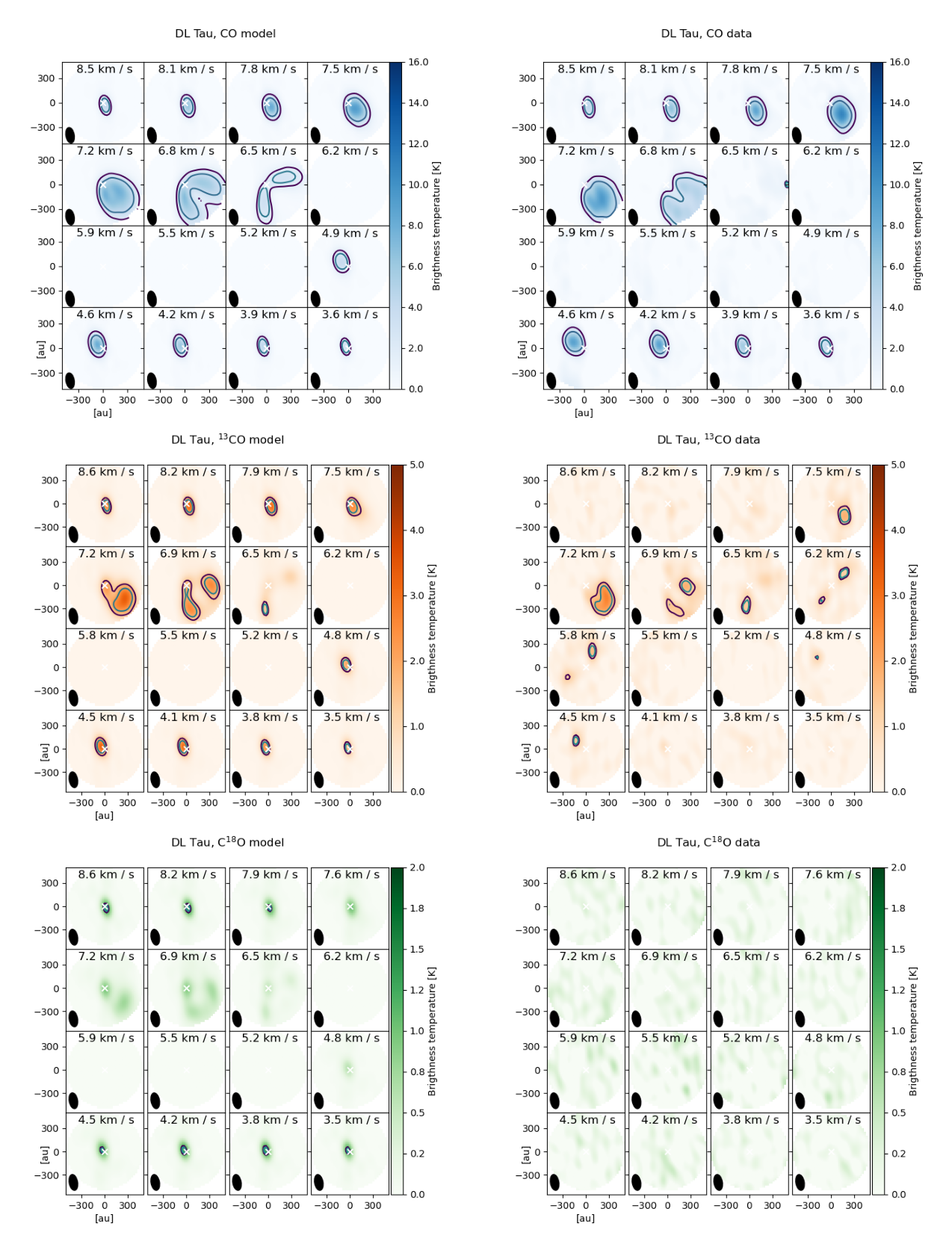}
     \caption{Comparison between the best-fit model and the observation of DL~Tau for the CO, $^{13}$CO, and C$^{18}$O isotopologue emission.}
     \label{fig: DL Tau images}
\end{figure*}

\begin{figure*}
\centering
   \includegraphics[height=0.9\textheight]{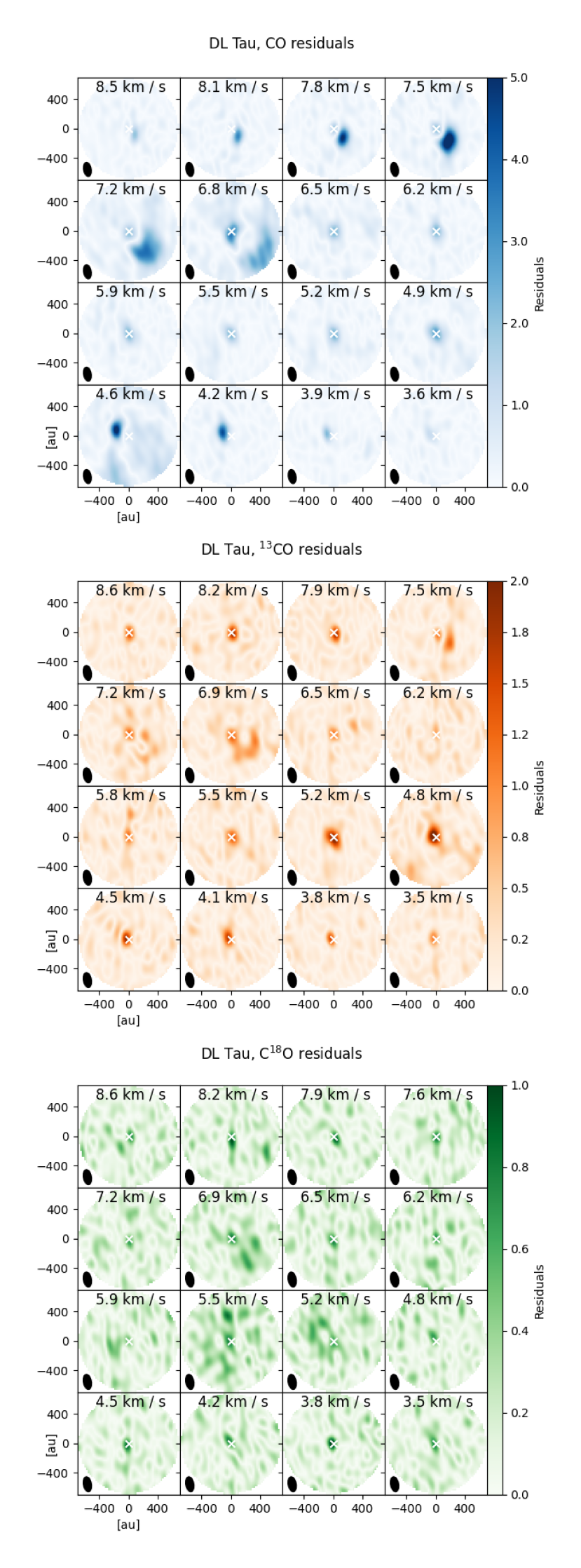}
     \caption{Residuals of DL~Tau best-fit model for the CO, $^{13}$CO, and C$^{18}$O isotopologue emission.}
     \label{fig: DL Tau images residuals}
\end{figure*}

    \begin{figure*}
\centering
   \includegraphics[height=0.9\textheight]{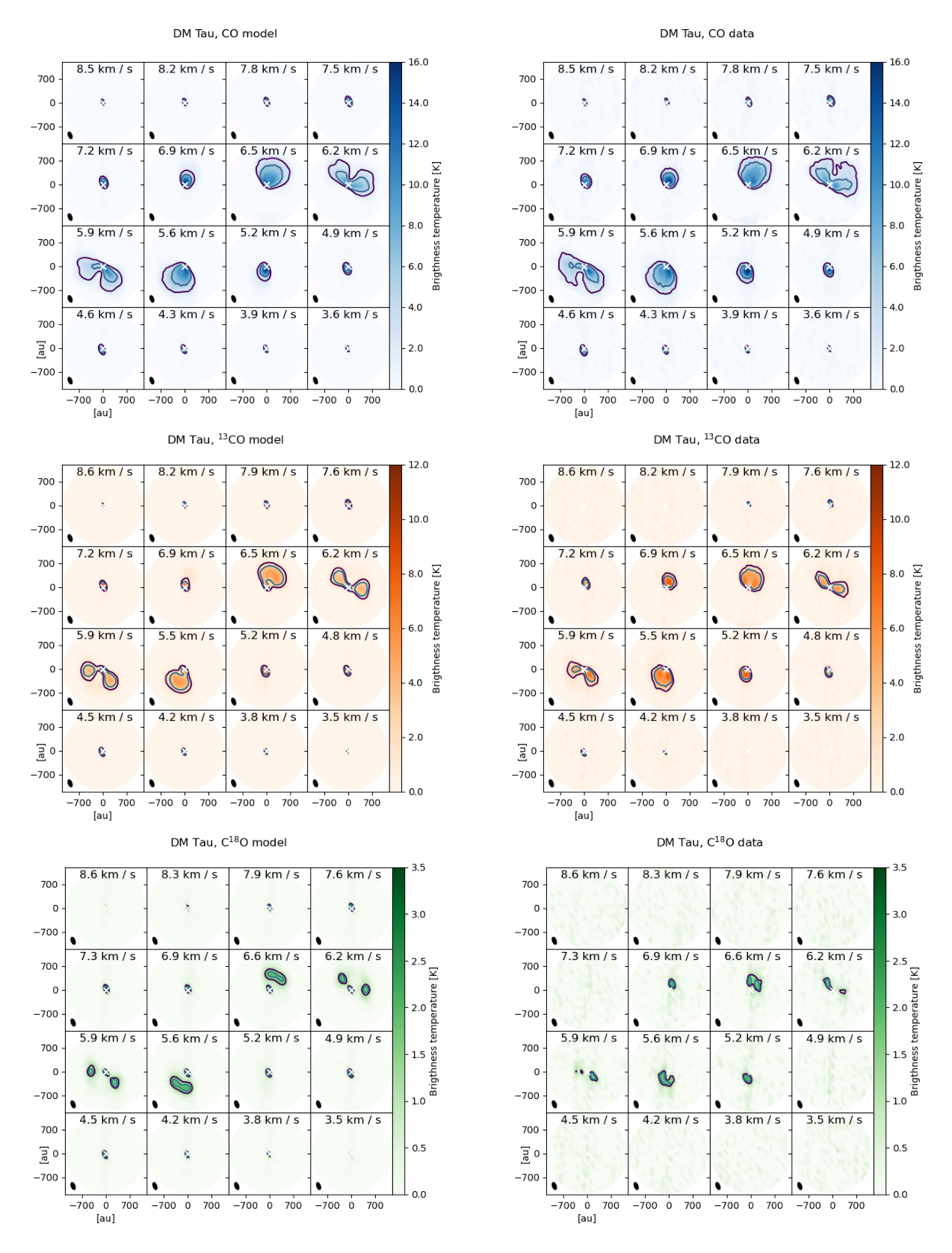}
     \caption{Comparison between the best-fit model and the observation of DM~Tau for the CO, $^{13}$CO, and C$^{18}$O isotopologue emission.}
     \label{fig: DM Tau images 2}
\end{figure*}

\begin{figure*}
\centering
   \includegraphics[height=0.9\textheight]{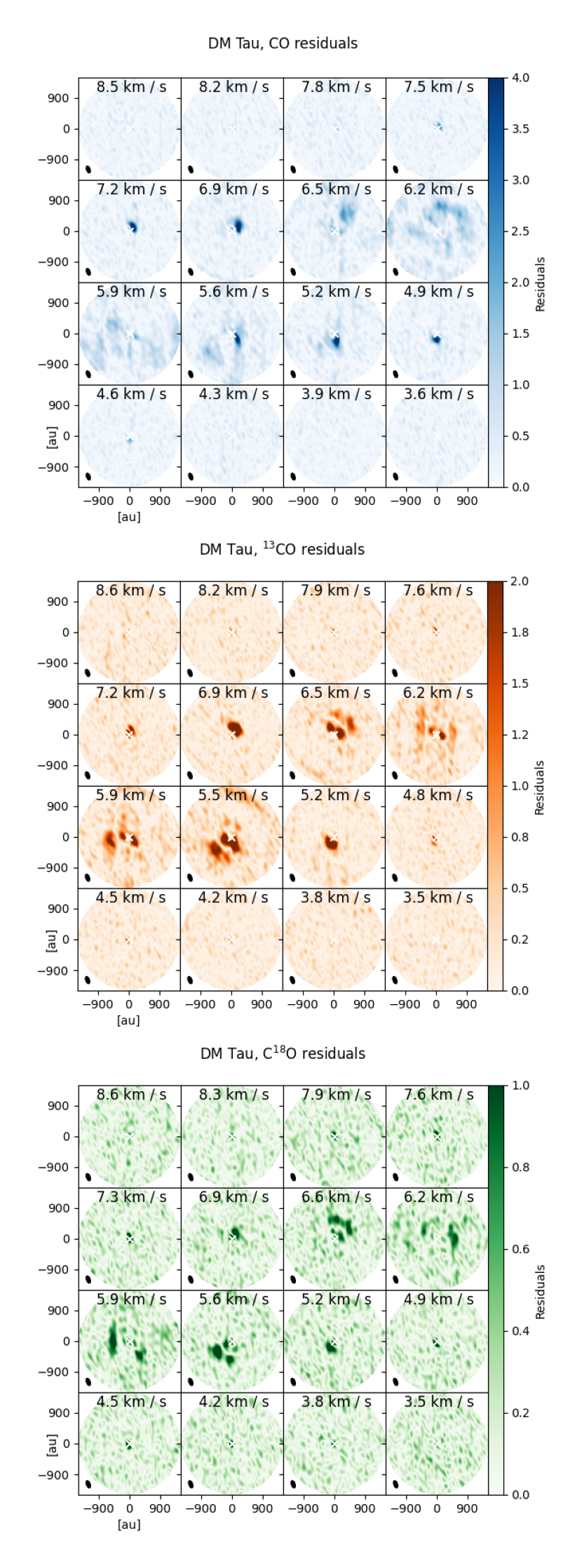}
     \caption{Residuals of DM~Tau best-fit model for the CO, $^{13}$CO, and C$^{18}$O isotopologue emission.}
     \label{fig: DM Tau images residuals 2}
\end{figure*}

    \begin{figure*}
\centering
   \includegraphics[height=0.9\textheight]{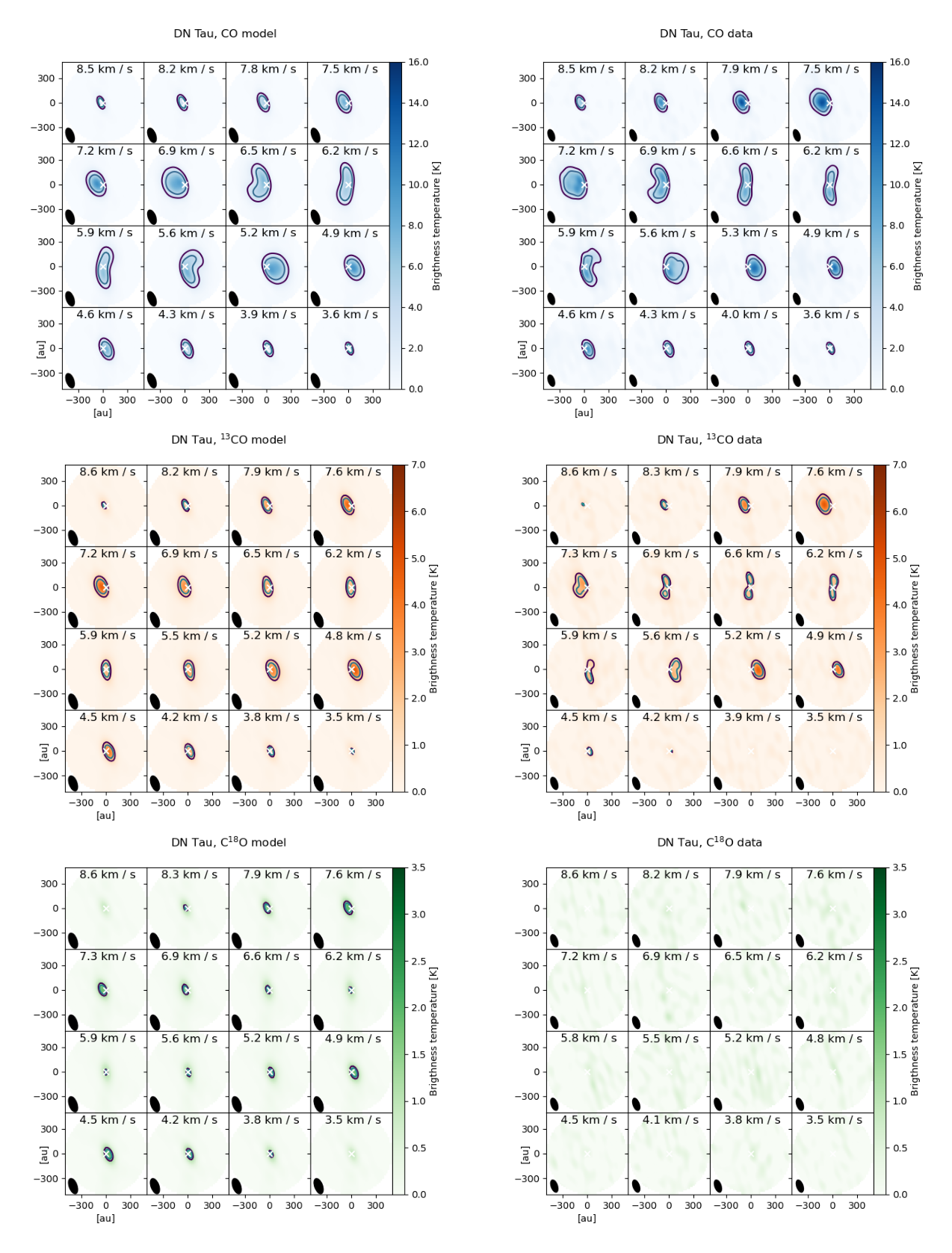}
     \caption{Comparison between the best-fit model and the observation of DN~Tau for the CO, $^{13}$CO, and C$^{18}$O isotopologue emission.}
     \label{fig: DN Tau images}
\end{figure*}

\begin{figure*}
\centering
   \includegraphics[height=0.9\textheight]{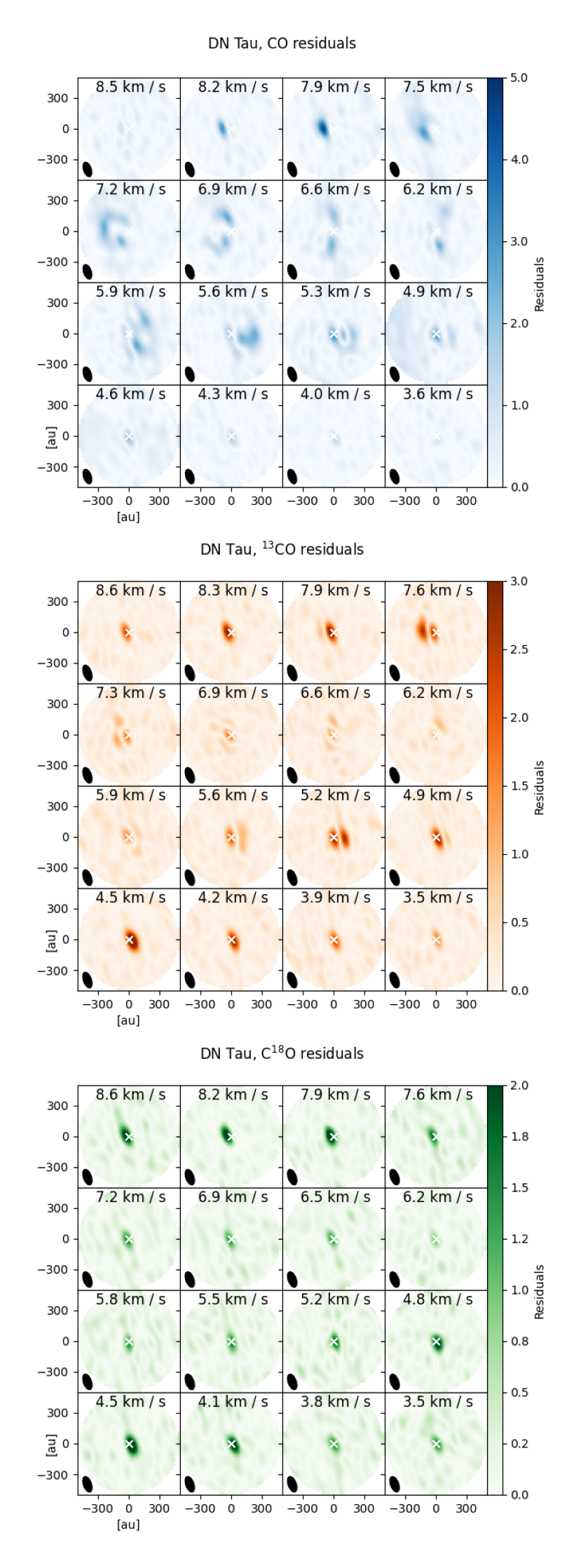}
     \caption{Residuals of DN~Tau best-fit model for the CO, $^{13}$CO, and C$^{18}$O isotopologue emission.}
     \label{fig: DN Tau images residuals}
\end{figure*}

    \begin{figure*}
\centering
   \includegraphics[height=0.9\textheight]{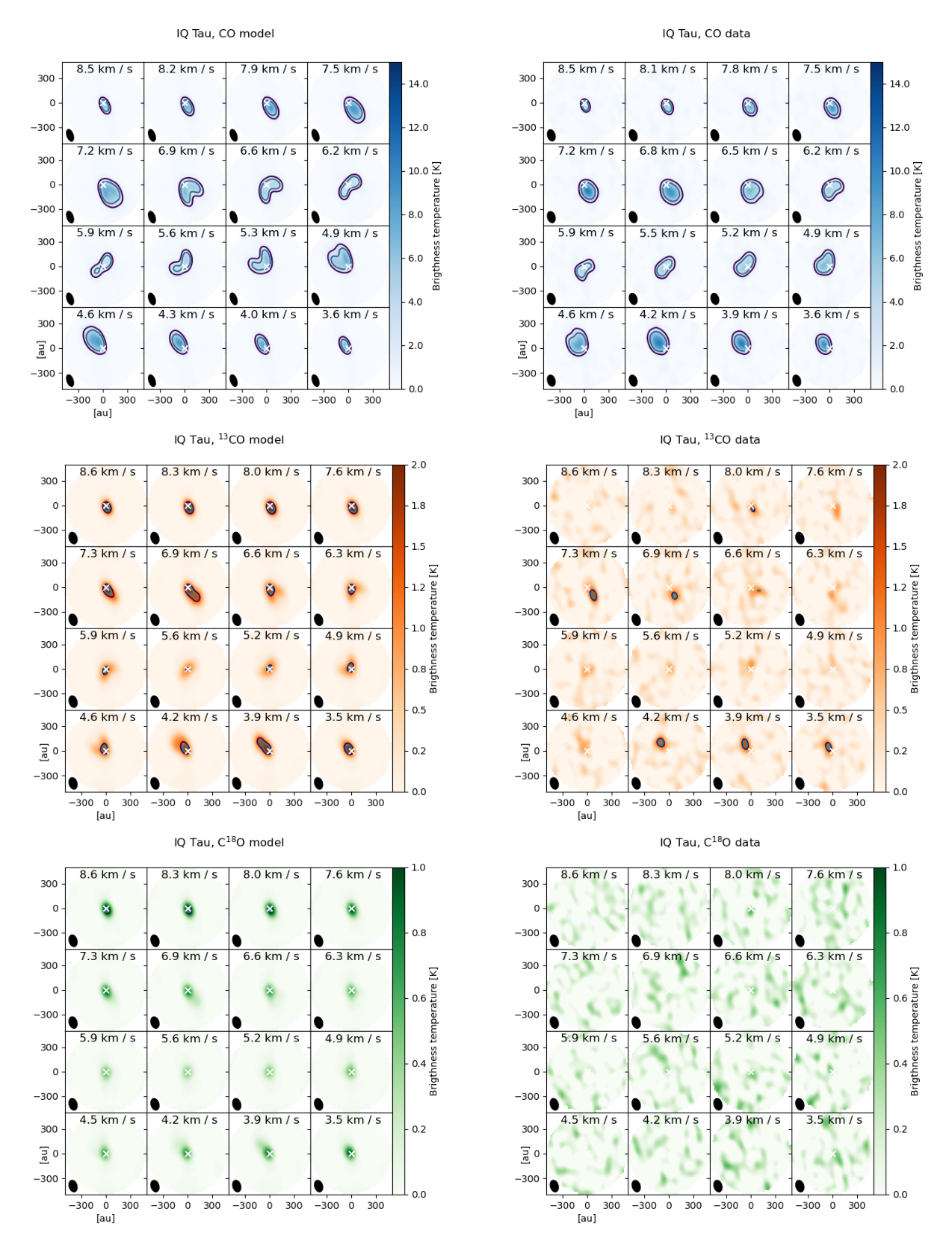}
     \caption{Comparison between the best-fit model and the observation of DM~Tau for the CO, $^{13}$CO, and C$^{18}$O isotopologue emission.}
     \label{fig: IQ Tau images}
\end{figure*}

\begin{figure*}
\centering
   \includegraphics[height=0.9\textheight]{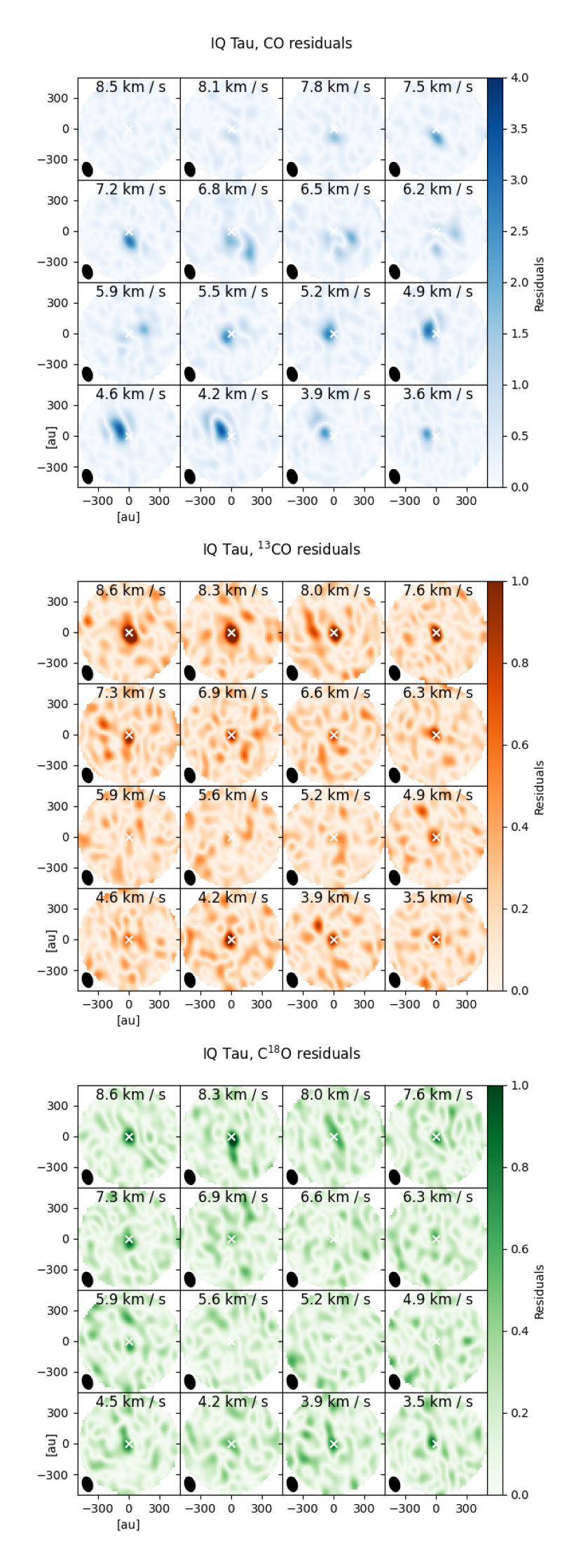}
     \caption{Residuals of IQ~Tau best-fit model for the CO, $^{13}$CO, and C$^{18}$O isotopologue emission.}
     \label{fig: IQ Tau images residuals}
\end{figure*}
\end{appendix}

\end{document}